\documentclass[a4paper,12pt]{article}         
\usepackage{geometry}           
\usepackage[page,titletoc,title]{appendix}
\usepackage{titlesec}
\usepackage{color}
\usepackage{cite}
\usepackage{latexsym,color}

\usepackage{epsfig}
\usepackage{amssymb,amsmath}
\usepackage{subcaption}
\usepackage[version=4]{mhchem}

\newcommand{\bc}{\begin{center}}
\newcommand{\ec}{\end{center}}
\newcommand{\bd}{\begin{displaymath}}
\newcommand{\ed}{\end{displaymath}}
\newcommand{\be}{\begin{equation}}
\newcommand{\ee}{\end{equation}}
\newcommand{\ba}{\begin{array}}
\newcommand{\ea}{\end{array}}
\newcommand{\bt}{\begin{tabular}}
\newcommand{\et}{\end{tabular}}

\usepackage{booktabs}   
\usepackage{mhchem}
\begin{document}
	
\title{Probing Dirac Dark Matter in Composite Higgs Models with Xenon and Argon targets}

\author{M.~G.~Belyakova\,,\quad R.~Nevzorov\\[5mm]
\itshape{I. E. Tamm Department of Theoretical Physics,}\\[0mm]
\itshape{Lebedev Physical Institute, Leninsky prospect 53, 119991 Moscow, Russia}}

\date{}

\maketitle
\begin{abstract}
\noindent
Recent advancements in direct detection (DD) experiments stimulate the investigation of the interactions
of dark matter (DM) with nucleons and nuclei. In the framework of Composite Higgs Models (CHMs)
the lightest Dirac composite particle (LDCP) can be stable composing a significant fraction $\xi$
of the observed DM relic abundance. We consider the elastic scattering of the LDCP on nucleons as well as
on xenon (Xe) and argon (Ar) nuclei within the CHMs in which the LDCP magnetic moment, its mass
and its coupling to the Higgs doublet are suppressed by an approximate $U(1)$ symmetry.
The LDCP with non--zero magnetic dipole moment can result in a substantial enhancement of the
differential event rate in the DD experiments at low recoil energies of nuclei. Assuming $\xi\ge 0.1$,
we identify the region of the parameter space where such enhancement may be potentially observable.
In addition we specify some observables that can be useful in discriminating between the DM fermions with
non--zero magnetic moment and other types of dark matter particles which don't have similar
electromagnetic properties.
\end{abstract}

\newpage
\section{Introduction}

Observations of merging galaxy clusters (such as the Bullet Cluster \cite{Dawson:2012fx, Randall:2008ppe}), galactic rotation
curves \cite{3.1, Bosma:2023kqn}, and gravitational lensing \cite{Massey:2010hh} provide compelling evidence
for the existence of dark matter (DM). At the same time
these astrophysical probes have not significantly advanced our knowledge of the fundamental properties of DM particles,
namely their mass, spin, electromagnetic couplings and other quantum numbers. To uncover these properties both direct and
indirect detection experiments have been conducted.

Direct detection experiments aim to measure nuclear recoil energies from elastic scattering of DM particles on nuclei \cite{Goodman:1984dc}.
Leading experiments, such as LUX--ZEPLIN (LZ) \cite{LZ:2024zvo}, Particle and Astrophysical Xenon Detector
(PandaX) \cite{PandaX:2024qfu}, XENONnT \cite{XENON:2025vwd} and DarkSide-20k \cite{DarkSide-20k:2025pbm}\footnote{The DarkSide-20k
dark matter detector is currently in its final construction and assembly phase at the Laboratori Nazionali del Gran Sasso.}
use noble gases (Xe or Ar) as detection media. Recently experiments LZ and PICO set stringent limits on the DM--nucleon
scattering cross section \cite{LZ:2024zvo} and on the DM magnetic moment \cite{PICO:2022ohk}.

Indirect detection experiments search for anomalous fluxes of stable Standard Model (SM) particles produced in DM annihilation
or decay. These anomalous fluxes may propagate from the regions with high DM density such as galactic centers.
However in bright galaxies distinguishing a DM signal from astrophysical backgrounds can be challenging.
Dwarf galaxies with their low radiation background are therefore worthwhile targets for DM searches. The Fermi Gamma--Ray Space
Telescope \cite{Fermi-LAT:2015att, Fermi-LAT:2016uux}, MAGIC \cite{MAGIC:2022acl}, VERITAS \cite{VERITAS:2024usn},
H.E.S.S. \cite{HESS:2022ygk} and HAWC \cite{HAWC:2023owv} have provided strong constraints on the annihilation cross section
for DM masses from a few GeV up to $10\,\mbox{TeV}$. The IceCube Neutrino Observatory \cite{IceCube:2023ies}, ANTARES \cite{ANTARES:2019svn},
the Baikal Deep Underwater Neutrino Telescope (BDUNT)\cite{Avrorin:2016yhw} and Super-Kamioka Neutrino Detection
experiment (SK) \cite{Super-Kamiokande:2020sgt}, which search for high-energy neutrino fluxes, set somewhat weaker limits
on this cross section. Using antiproton data from AMS-02 \cite{AMS:2016oqu} a stringent bound on the DM annihilation cross section
was also obtained in \cite{Cuoco:2017iax}.

An attractive DM candidate appears in supersymmetric (SUSY) extensions of the SM (for a recent
review see \cite{Nevzorov:2023dhd}). If R--parity is conserved
the lightest neutralino can be stable and contribute to the DM density. Being heavy weakly interacting massive particles (WIMPs)
such neutralinos were non-relativistic in the early Universe and could gather together before the recombination stage helping
to explain the large--scale structure formation of the Universe\cite{Primack:2002th}.
The gauge coupling unification in the minimal supersymmetric standard model (MSSM) permits to embed MSSM into
Grand Unified Theories (GUTs) \cite{Georgi:1974sy} based on simple gauge groups such as $SU(5)$, $SO(10)$ or $E_6$.
On the other hand the cancellation of quadratic divergences \cite{mass-hierarchy} in models with softly broken SUSY allows
to almost stabilize the electroweak (EW) scale solving the hierarchy problem \cite{Gildener:1976ai}.
Nevertheless, scenarios with a stable lightest neutralino are strongly constrained by direct and indirect searches.
Also LHC experiments have already excluded some part of the parameter space of SUSY models.
	
In this context it is worth to examine alternative DM scenarios within other well motivated extensions of the SM.
Composite Higgs models (CHMs), for instance, can provide suitable DM candidates. Strong dynamics in these models
is expected to give rise to a set of resonances with quantum numbers of the SM particles including the Higgs boson.
The same dynamics may also yield a neutral Dirac fermion $\chi$ that composes the DM. In our previous article \cite{Belyakova:2024fcw}
we explored such a scenario within the $E_6$ inspired composite Higgs model (E$_6$CHM) \cite{Nevzorov:2015sha}--\cite{Nevzorov:2016fxp}
with an approximate $U(1)_E$ symmetry. Within the E$_6$CHM the Higgs doublet emerges as a set of pseudo-Nambu-–Goldstone (pNG) bosons
from the spontaneous breaking of an approximate $SU(6)$ symmetry of the strongly coupled sector down to its $SU(5)$ subgroup
near the scale $f\gtrsim 5\,\mbox{TeV}$. Conservation of baryon number $B$ ensures that the lightest, $SU(5)$ singlet, composite fermion
$\chi$ with $B=1/3$ is stable in this model.

The Dirac DM fermion $\chi$ generally possesses a non--zero magnetic dipole moment $\mu_{\chi}$, which is strongly
constrained \cite{PICO:2022ohk}\footnote{The presence of very light neutral fermions with non--zero magnetic moment
in the particle spectrum might have interesting implications for the neutrino physics \cite{Frere:1996gb}.}.
The corresponding term in the E$_6$CHM Lagrangian violates the approximate $U(1)_E$ symmetry
so that $\mu_{\chi}$ is expected to be suppressed. In our previous analysis we assumed $\mu_{\chi}$ to be negligibly small \cite{Belyakova:2024fcw}
and examined constraints on the E$_6$CHM parameter space caused by the stringent bounds on the spin--independent
DM--nucleon scattering cross section. We found phenomenologically viable regions of the parameter space associated with
$f\simeq 5-10\,\mbox{TeV}$ and $m_{\chi}\gtrsim 200\,\mbox{GeV}$, which may also yield spectacular LHC signatures.

The goal of this paper is to explore the interactions of $\chi$ with xenon (Xe) and argon (Ar) nuclei taking into account
the non--zero value of $\mu_{\chi}$ and assuming CP conservation in the strongly interacting sector so that
the DM electric dipole moment can be ignored.
Electromagnetic interaction of spin--$1/2$ DM particles, which possess the magnetic
dipole moment, with nucleons and nuclei were considered before \cite{Ibarra:2024mpq,Banks:2010eh,Hambye:2021xvd}.
Nonetheless in the CHMs the Dirac DM fermion $\chi$ interacts with nuclei not only through long--range electromagnetic force but
also via short--range interactions caused by the $t$-channel exchanges of the $Z$ boson and the SM Higgs scalar.
Here we limit our consideration by those regions of the parameter space of the CHMs
where $\chi$ can account for a significant fraction $\xi$ of the total DM density $\rho_{DM}$, i.e. $\xi \gtrsim 0.1$.
In our analysis the compositeness scale $f$ is varied from $5\,\mbox{TeV}$ to $15\,\mbox{TeV}$.
We investigate the dependence of the differential $\chi$--Xe (Ar) scattering cross section on the recoil energies of nuclei,
focusing on the parameter regions, where $\mu_{\chi}$ can result in substantial enhancement of this cross section.
The normalized differential event rate is studied and the fraction of the events with low recoil energies of nuclei is computed.
We also discuss observables that may permit to differentiate between DM states with $\mu_{\chi}\ne 0$ and
$\mu_{\chi}= 0$.

The layout of this article is as follows. In Section~2 we briefly review composite Higgs models
and specify the couplings of the Dirac DM fermions. Using these couplings in Section~3 the spin--independent and spin--dependent
DM--nucleon scattering cross sections are calculated. These cross sections are compared with the current experimental bounds
and the regions of the parameter space, where the Dirac DM fermions $\chi$ can comprise a substantial fraction of DM,
are identified. The results of our numerical analysis of the differential cross section of $\chi$--Xe and $\chi$--Ar elastic
scattering are presented in Section~4. The corresponding normalized differential event rates are also studied and
the appropriate observables are considered. Section~5 concludes the paper.
	
\section{Composite Higgs models and Dirac DM fermions}
	
Extensions of the SM, which include strongly coupled sector, that leads to the composite Higgs doublet,
are called composite Higgs models (CHMs) (for a review, see \cite{Bellazzini:2014yua}).
The minimal composite Higgs Model (MCHM) \cite{Agashe:2004rs} possesses an approximate $\mbox{SO(5)}\times U(1)_X$ symmetry which contains the
$SU(2)_W\times U(1)_Y$ gauge group as a subgroup. Under this symmetry, the Higgs doublet transforms as a set of pseudo-Nambu-Goldstone bosons (pNGBs),
much like the pions in QCD. Around the scale $f$ this approximate symmetry gets spontaneously broken down to
$SO(4)\times U(1)_X \cong SU(2)_W\times SU(2)_R\times U(1)_X $, so that $SU(2)_W\times U(1)_Y$ and custodial symmetry
$SU(2)_{cust} \subset SU(2)_W\times SU(2)_R$ remain intact. In general the compositeness scale $f$ has to be larger than $10\,\mbox{TeV}$,
because of the constraints which stem from the measurements of the electron electric dipole moment and $\mu\to e\gamma$ transitions \cite{Agashe:2006iy}
as well as the measurements of CP violation in the Kaon system \cite{Csaki:2008zd}--\cite{Barbieri:2012tu}.
This bound can be substantially alleviated in the CHMs with additional flavor symmetries \cite{Barbieri:2012tu}--\cite{Cacciapaglia:2007fw},
so that even the scenarios with $f\sim 1\,\mbox{TeV}$ can be phenomenologically viable. For $f\ll 10\,\mbox{TeV}$
one also needs to impose global $U(1)_B$ symmetry associated with the baryon number conservation to suppress
neutron--antineutron oscillations as well as rapid proton decay.
	
Different extensions of the MCHM were discussed in Refs. \cite{Belyakova:2024fcw}--\cite{Nevzorov:2016fxp}, \cite{Cacciapaglia:2014uja}--\cite{Chala:2016ykx}. The implications of these models were explored for
dark matter \cite{Belyakova:2024fcw,Frigerio:2011zg,Chala:2016ykx}, baryogenesis \cite{Nevzorov:2017rtf,Chala:2016ykx}
and leptogenesis \cite{Nevzorov:2025ido}. Strong dynamics in these models can give rise to two SM singlet Weyl fermions
which compose the neutral Dirac state $\chi$ with reduced couplings to SM fields. If $\chi$ is stable it may form the DM in our Universe.
	
As an example, let us consider the E$_6$CHM \cite{Nevzorov:2015sha}\footnote{The $E_6$ inspired supersymmetric extensions
of the SM were studied in \cite{e6ssm}.}. The weakly--coupled sector of the E$_6$CHM
involves the following set of elementary states
\begin{equation}
(q_i,\,d^c_i,\,\ell_i,\,e^c_i) + u^c_{\alpha} + \bar{q}+\bar{d^c}+\bar{\ell}+\bar{e^c}\,,
\label{1}
\end{equation}
where $\alpha=1,2$ and $i=1,2,3$. In Eq.~(\ref{1}) we have denoted
the SM right-handed up- and down-type quarks and charged leptons by $u_{\alpha}^c, d_i^c$ and $e_i^c$,
the SM left-handed quark and lepton doublets by $q_i$ and $\ell_i$, whereas exotic states $\bar{q},\,\bar{d^c},\,\bar{\ell}$
and $\bar{e^c}$ have exactly opposite $SU(3)_C\times SU(2)_W\times U(1)_Y$ quantum numbers to
left-handed quark doublets, right-handed down-type quarks, left-handed lepton doublets and
right-handed charged leptons. Since the strongly interacting sector of the E$_6$CHM possesses the approximate $SU(6)$ symmetry
all composite resonances in this model belong to complete $SU(6)$ representation. In particular, it is expected that
the dynamics of the strongly coupled sector gives rise to one ${\bf{15}}$--plet and two ${\bf \overline{6}}$--plets
(${\bf \overline{6}}_1$ and ${\bf \overline{6}}_2$) of $SU(6)$ which decompose under $SU(3)_C\times SU(2)_W\times U(1)_Y\times U(1)_B$
as follows:
\begin{equation}
\ba{ll}
\ba{rcl}
{\bf 15} &\to& Q = \left(3,\,2,\,\dfrac{1}{6},\,-\dfrac{1}{3}\right)\,,\\[2mm]
&& t^c = \left(\bar{3},\,1,\,-\dfrac{2}{3},\,-\dfrac{1}{3}\right)\,,\\[2mm]
&& E^c = \Biggl(1,\,1,\,1,\,-\dfrac{1}{3}\Biggr)\,,\\[2mm]
&& D = \left(3,\,1,\,-\dfrac{1}{3},\,-\dfrac{1}{3} \right)\,,\\[2mm]
&& \overline{L}=\left(1,\,2,\,\dfrac{1}{2},-\dfrac{1}{3}\,\right)\,;
\ea
\qquad
\noindent
\ba{rcl}
{\bf \overline{6}}_{1} &\to & D^c_{1} = \left(\bar{3},\,1,\,\dfrac{1}{3},\,\dfrac{1}{3} \right)\,,\\[2mm]
& & L_{1} = \left(1,\,2,\,-\dfrac{1}{2},\,\dfrac{1}{3} \right)\,,\\[2mm]
& & N_{1} = \Biggl(1,\,1,\,0,\,\dfrac{1}{3} \Biggr)\,;\\[6mm]
{\bf \overline{6}}_{2} &\to & D^c_{2} = \left(\bar{3},\,1,\,\dfrac{1}{3},\,-\dfrac{1}{3} \right)\,,\\[2mm]
& & L_{2} = \left(1,\,2,\,-\dfrac{1}{2},\,-\dfrac{1}{3} \right)\,,\\[2mm]
& & \overline{N}_{2} = \Biggl(1,\,1,\,0,\,-\dfrac{1}{3} \Biggr)\,.
\ea
\ea
\label{2}
\end{equation}
Here the first and second quantities in brackets are the $SU(3)_C$ and $SU(2)_W$ representations, while the third
and fourth quantities are $U(1)_Y$ and $U(1)_{B}$ charges.
	
The breakdown of $SU(6)$ down to $SU(5)$, that includes the $SU(3)_C\times SU(2)_W\times U(1)_Y$ subgroup, results in
a set of the following mass terms
\be
\mathcal{L}_{mass} = \mu_{q} \bar{q} Q + \mu_e \bar{e^c} E^c + \mu_{D} D^c_{1} D + \mu_{L} \overline{L} L_{1}
+ \mu_d \bar{d^c} D^c_{2} + \mu_{l} \bar{\ell} L_{2} + \mu_N \overline{N}_{2} N_{1} + h.c.\,.
\label{3}
\ee
In general all mass parameters $\mu_{I}$ in Eq.~(3) are of order of $f$.
Because $SU(5)$ does not contain $SU(2)_{cust}$ subgroup, the electroweak precision measurements
set lower bound $f\gtrsim 5-6\,\mbox{TeV}$ in the E$_6$CHM \cite{Nevzorov:2015sha}.
According to Eq.~(\ref{3}) $\bar{q},\,\bar{d^c},\,\bar{\ell}$, $\bar{e^c}$ as well as
all components of ${\bf \overline{6}}_1$, ${\bf \overline{6}}_2$ and ${\bf{15}}$ except $t^c$ gain
large masses. Only composite right--handed top quark $t^c$ survives to the electroweak (EW) scale.
	
The $SU(6)$ symmetry breaking down to its $SU(5)$ subgroup leads to eleven pNGB states which form
the scalar colour triplet $T$, the SM Higgs doublet $H$ and the SM singlet pseudoscalar $\phi_0$.
All pNGBs do not carry any baryon and/or lepton numbers and their masses tend to be considerably
smaller than $f$. Then the baryon number conservation in the E$_6$CHM ensures that the colour triplet $T$
is stable. Such scenarios have been already ruled out. The composite Higgs model under consideration
is phenomenologically viable only if the E$_6$CHM Lagrangian is invariant under the transformations of
an approximate $U(1)_E$ symmetry defined as
\be
{\bf \overline{6}_2} \longrightarrow e^{i\beta} {\bf \overline{6}_2},\qquad \bar{d^c} \longrightarrow e^{-i\beta} \bar{d^c},\qquad
\bar{\ell} \longrightarrow e^{-i\beta} \bar{\ell}\,.
\label{4}
\ee
From Eq.~(4) it follows that the contributions of exotic fermions to $U(1)_E$ - $SU(3)_C$ - $SU(3)_C$,
$U(1)_E$ - $SU(2)_W$ - $SU(2)_W$ and $U(1)_E$ - $U(1)_Y$ - $U(1)_Y$ anomalies get cancelled.
If $U(1)_E$ were exact $\mu_N$ would vanish. Thus the approximate $U(1)_E$ symmetry yields $\mu_N\ll f$ so that
the lightest composite fermion $\chi$ in the spectrum is mostly a superposition of $N_{1}$ and $N_{2}$\footnote{It
is assumed that the violation of the $U(1)_E$ global symmetry does not lead to any CP violating terms.}.
The fermion $\chi$ carries baryon number $B=1/3$ and cannot decay into SM particles \cite{Belyakova:2024fcw}.
It can be substantially lighter than the scalar colour triplet $T$. As a consequence $T$ can decay into $b$--quark
and $\overline{\chi}$. At the LHC the pairs of $T\overline{T}$ are mostly produced through the gluon fusion resulting
in some enhancement of the cross section of
\begin{equation}
pp\to b\overline{b} + E^{\rm miss}_{T} + X\,,
\label{5}
\end{equation}
where $E^{\rm miss}_{T}$ is the missing energy and transverse momentum associated with $\chi$ and $\overline{\chi}$ in the final state.

The stability of the lightest composite fermion $\chi$ is caused by the global $U(1)_{B}$ symmetry
which is imposed in the CHMs to suppress the baryon number violating interactions. Thus, like in SUSY models, dark matter
arises in the E$_6$CHM as a consequence of solving proton stability.
Moreover there is a discrete $Z_3$ symmetry, kind of analogous to $R$--parity in supersymmetry,
which is known as baryon triality \cite{baryon-triality}. The corresponding symmetry transformations are given by
\begin{equation}
\Psi \longrightarrow e^{2\pi i B_3/3} \Psi,\qquad B_3 = (3 B - n_C)_{\mbox{mod}\,\, 3}\,,
\label{51}
\end{equation}
where $B$ is the baryon number of the multiplet $\Psi$ and $n_C$ is the number of colour indices ($n_C=1$ for the
colour triplet and $n_C=-1$ for antitriplet). The baryon number conservation implies that the E$_6$CHM Lagrangian is also invariant
under the symmetry transformations (\ref{51}). All SM bosons and SM fermions have $B_3=0$ while exotic fermions
and the scalar colour triplet $T$ carry either $B_3=2$ or $B_3=1$. Therefore the lightest composite fermion $\chi$
with non--zero $B_3$ charge cannot decay into SM particles and should be stable.
	
The lightest Dirac composite particle $\chi$ (LDCP) can form the DM in our Universe.
To ensure the phenomenological viability of the scenario under consideration we assume
that the mass of the lightest exotic fermion $m_{\chi}$ is quite close to half the mass
of the SM singlet pNGB state $\phi_0$, i.e. $m_A/2$ \cite{Belyakova:2024fcw}.
In the Yukawa interactions with fermions $\phi_0$ manifests itself as a pseudoscalar field.
When $m_{\chi}\approx m_A/2$ the annihilation cross section for $\chi+\bar{\chi}\to t + \bar{t}$
can be large enough to ensure that the LDCP accounts for all or some of the observed
cold dark matter density.
	
Hereafter we just assume that two SM singlet Weyl fermions $N_1$ and $N_2$ form the neutral LDCP $\chi$ ($\chi_L=N_1$ and $\chi_R=N_2$)
which contributes to the DM relic abundance. The corresponding contribution to the total DM density is defined by some unknown mechanism.
If such LDCP exists its couplings to the SM particles are strongly suppressed. For instance, the coupling of the LDCP to
the Z--boson is determined by the following non-renormalizable operator:
\begin{equation}
\mathcal{L}_Z=\frac{\lambda_1}{f^2}H^{+}iD_{\mu}H\bar{N}_1\gamma_{\mu}N_1+\frac{\lambda_2}{f^2}H^{+}iD_{\mu}H\bar{N}_2\gamma_{\mu}N_2 \qquad
\label{6}
\end{equation}
which can be induced by the non-perturbative dynamics in the CHMs. In Eq.~(\ref{6})
$\lambda_1$ and $\lambda_2$ are dimensionless couplings of order unity. Operators in Eq.~(\ref{6}) lead to
\begin{equation}
\mathcal{L}_{Z\chi\chi}=\overline{\chi} (a^{\chi}_{V} \gamma^{\mu} + a^{\chi}_{PV} \gamma^{\mu} \gamma^{5})\chi Z_{\mu},
\label{7}	
\end{equation}
\begin{equation}
a^{\chi}_{V} = \dfrac{\bar{g} \eta^2}{8 f^2}c_V^{\chi} \qquad 	a^{\chi}_{PV}=\dfrac{\bar{g} \eta^2}{8 f^2}c_{PV}^{\chi},
\label{8}
\end{equation}
where  $c_V^{\chi}=\lambda_1+\lambda_2$, $c_{PV}^{\chi}=\lambda_1-\lambda_2$, $\bar{g}=\sqrt{g^2+g'^2}$, $g$ and $g'$ are $SU(2)_W$ and $U(1)_Y$
coupling constants correspondingly and $\eta\simeq 246\,\mbox{GeV}$ is a vacuum expectation value (VEV) of the Higgs field.

The interactions of $\chi$ with the Higgs doublet $H$ and electromagnetic field are given by
\begin{equation}
\mathcal{L_{\chi H}}=\frac{\varepsilon_H}{f} H^{\dagger} H (\overline{N}_{2} N_{1}) + h.c,
\label{9}
\end{equation}
\begin{equation}
\mathcal{L_{\chi M}}=\frac{\mu_{\chi}}{2}\overline{N}_2\sigma^{\mu\nu} N_1 F_{\mu\nu} +h.c.\,.
\label{10}
\end{equation}
Here $F_{\mu\nu}=\partial_{\mu}A_{\nu}-\partial_{\nu}A_{\mu}$ and $A_{\mu}$ is electromagnetic field.
In the CHMs it is expected that $\mu_{\chi}\sim e/f$ where $e$ is the electron charge.
On the other hand the DM direct detection experiment sets stringent upper bound on the magnetic dipole moment of DM fermions,
i.e. $|\mu_{\chi}|\lesssim 10^{-8}\,\mbox{GeV}$ \cite{PICO:2022ohk}. Naively, it looks like so stringent
experimental limit on $|\mu_{\chi}|$ ruled out CHMs with $f\lesssim 10^4\,\mbox{TeV}$ if DM is accounted for by
Dirac fermions. However the DM magnetic dipole moment in these models tends to be much smaller than $e/f$ if
all constituents, which form the DM composite fermion, have zero electric charges. In this case non--zero value of $\mu_{\chi}$
is induced via triangle loop diagrams\footnote{These one--loop Feynman diagrams involve charged fermions from
${\bf \overline{6}}_1$ and ${\bf \overline{6}}_2$ as well as charged scalar and vector bosons from adjoint representations of $SU(6)$.}.
When $m_{\chi}\ll f$ the computation of these diagrams is rather similar to the calculation
of the magnetic moment of the Dirac neutrino in the SM. Then the magnetic moment of the DM composite fermions
can be estimated as
\begin{equation}
|\mu_{\chi}|\simeq \dfrac{c_{\mu}}{16\pi^2} \left(\dfrac{m_{\chi}}{f}\right) \left(\dfrac{e}{f}\right)\,,\qquad\qquad  c_{\mu}\sim 1\,,
\label{101}
\end{equation}
where the dimensionless coefficient $c_{\mu}$ is proportional to the couplings of exotic fermion
$\chi$ to other composite states and depends on the ratios of their masses. Because here we focus on the scenarios with
$m_{\chi}\lesssim 1\,\mbox{TeV}$ the experimental constraint on $\mu_{\chi}$ can be satisfied even for $f\sim 10\,\mbox{TeV}$.
Hereafter we assume that like in the E$_6$CHM the interactions (\ref{9})--(\ref{10}) are suppressed by some global
$U(1)$ symmetry, i.e. $\varepsilon_H\ll 1$ and $\mu_{\chi}\ll \dfrac{e}{f}$.

For further analysis  of $\chi$-nucleus scattering, we need to define the relevant nucleon coupling constants to
the Z--boson, Higgs scalar and photon. The interactions of quarks with the Z--boson can be written as
\begin{equation}
\begin{array}{c}
\mathcal{L}_{Zq}=\sum_{q}\dfrac{\bar{g}}{2}\bar{q}\gamma^{\mu}\left(a_V^q+a_{PV}^q\gamma_5\right)qZ_{\mu}=\dfrac{\bar{g}}{2}J_{ NC}^{\mu}Z_{\mu},\\
a_V^q=\left(T_3^q-2Q_qs^2_W\right),\qquad  a_{PV}^q=T_3^q.
\end{array}
\label{11}
\end{equation}
where $s_W=\dfrac{g'}{\bar{g}}$, $Q_q$ and $T_3^q$ are the electric charge and the third component of isospin of quark $q$.
At low energies, the corresponding matrix elements for nucleons $N$ are given by
\begin{equation}
\langle N'|J_{ NC}^{\mu}|N \rangle=\bar{\psi}'_N\gamma^{\mu}\left(a_V^N+a_{PV}^N\gamma_5\right)\psi_N\,,
\label{12}
\end{equation}
\begin{equation}
a_V^p\simeq\left(\frac{1}{2}-2s^2_W\right),\qquad
a_{PV}^p\simeq\left(\frac{1}{2}\Delta^{(p)}_u-\frac{1}{2}\Delta^{(p)}_d-\frac{1}{2}\Delta^{(p)}_s\right)\,,
\label{13}
\end{equation}
\begin{equation}
a_V^n\simeq -\frac{1}{2},\qquad
a_{PV}^n\simeq \left(\frac{1}{2}\Delta^{(n)}_u-\frac{1}{2}\Delta^{(n)}_d-\frac{1}{2}\Delta^{(n)}_s\right)\,.
\label{14}
\end{equation}
Using $\Delta^{p}_u=\Delta^{n}_d=0.777$, $\Delta^{p}_d=\Delta^{n}_u=-0.438$ and $\Delta^{p}_s=\Delta^{n}_s=-0.053$ \cite{Lin:2018obj}
one obtains $a_{PV}^p \simeq 0.63$ and $a_{PV}^n \simeq -0.58$.

The interactions of quarks with the Higgs field $h$ can be presented in the following form
\begin{equation}
\mathcal{L}_{Hq}=\sum_{q}\frac{m_q}{\eta}(\bar{q}q) h\,,
\label{15}
\end{equation}
where  $m_q$ are the masses of $u,\,d,\,s,\,c,\,b$ and $t$ quarks.
The corresponding nucleon matrix element can be parameterized as
\begin{equation}
\frac{1}{\eta}\langle N'|\sum_{q}m_q\bar{q}q |N\rangle=a_H^N\frac{m_N}{\eta}\bar{\psi}'_N\psi_N\,.
\label{eqc.0}
\end{equation}
where $m_N$ is a nucleon mass and
$$
a_H^N = \sum_{q} f^N_{Tq}\,,\qquad m_N f^N_{Tq}=\langle N | m_{q}\bar{q}q |N \rangle\,.
$$
Using the results of calculations of $f^N_{Tq}$ \cite{Ellis:2018dmb} one finds $a_H^N\simeq 0.29$.

The nucleon electromagnetic current can be expressed as
\begin{equation}
\begin{array}{rcl}
j^N_{\mu}&=&\left<N'\right|\sum_q e_q \bar{q}\gamma_{\mu}q\left|N\right>=\\
&=&\bar{\psi}'_{N}(k')\left[ \left(4m_N^2F_e^N-q^2F_m^N\right)\dfrac{\gamma_{\mu}}{P^2}
+\dfrac{2m_N}{P^2}\left(F_e^N-F_m^N\right)\sigma_{\mu\nu}q^{\nu}\right]\psi_{N}(k),
\end{array}
\label{17}
\end{equation}
where $q^{\nu}=k'^{\nu}-k^{\nu}$,  $P^{\nu}=k'^{\nu}+k^{\nu}$, $F_e^N$ and $F_m^N$ are the electric and magnetic
nucleon form factors which depend on $q^2$. When $q^2$ vanishes $F_e^p \simeq 1$, $F_e^n \simeq 0$, $F^p_m\simeq 2.8$ and $F^n_m\simeq -1.9$.

\section{Interaction of Dirac DM fermions with nucleons}

The explicit expressions (\ref{7})--(\ref{17}) for the interactions of the neutral gauge bosons and the Higgs particle with fermions
permit to compute the $\chi$--nucleon scattering amplitude $\mathcal{M}_{\chi N}$. In the CHMs
the $t$-channel exchanges of the $Z$ boson, photon and the SM Higgs scalar contribute to this amplitude.
In the non--relativistic limit one gets
\begin{equation}
\mathcal{M}_{\chi N}\simeq \Biggl(2 e m_N \mu_{\chi} F_e^N + \dfrac{4 m_{\chi} m^2_N}{m^2_H f} \varepsilon_H a_H^N
- \dfrac{2 \bar{g} m_{\chi} m_{N}}{m^2_Z} a^{\chi}_{V} a_V^N \Biggr)\mathcal{O}_1
\label{18}
\end{equation}
$$
+ \dfrac{8 e m_N m_{\chi} \mu_{\chi}}{|{\bf q}|^2} F_e^N \mathcal{O}_5
+\Biggl(\dfrac{8 \bar{g} m_{\chi} m_{N}}{m^2_Z} a^{\chi}_{PV} a_{PV}^N + 8 e m_{\chi} \mu_{\chi} F_m^N\Biggr)\mathcal{O}_4
- \dfrac{8 e m_{\chi} \mu_{\chi}}{|{\bf q}|^2} F_m^N \mathcal{O}_6\,,
$$
where
$$
\mathcal{O}_1 = (\xi^{'\dagger}_{\chi} \xi_{\chi})(\xi^{'\dagger}_{N} \xi_{N})\,,\qquad
\mathcal{O}_5 = i \left((\xi^{'\dagger}_{\chi} \hat{{\bf S}}_{\chi} \xi_{\chi})\cdot [{\bf q}\times {\bf v}_{\perp}]\right)
(\xi^{'\dagger}_{N} \xi_{N})\,,
$$
\begin{equation}
\mathcal{O}_4= (\xi^{'\dagger}_{\chi} \hat{{\bf S}}_{\chi} \xi_{\chi})\cdot (\xi^{'\dagger}_{N} \hat{{\bf s}}_{N} \xi_{N})\,,\qquad
\mathcal{O}_6= \left((\xi^{'\dagger}_{\chi} \hat{{\bf S}}_{\chi} \xi_{\chi})\cdot {\bf q}\right)
\left((\xi^{'\dagger}_{N} \hat{{\bf s}}_{N} \xi_{N})\cdot {\bf q}\right)\,,
\label{19}
\end{equation}
$$
{\bf v}_{\perp}={\bf v}_{\chi} - {\bf v}_{N} - \dfrac{{\bf q}}{2 m_r}\,,\qquad m_r=\dfrac{m_{\chi} m_N}{m_{\chi}+m_N}\,.
$$
Here $m_Z$ and $m_H$ are the masses of the $Z$--boson and Higgs scalar, i.e. $m_Z \simeq 91.2\,\mbox{GeV}$ and $m_H \simeq 125\,\mbox{GeV}$.
In Eqs.~(\ref{19}) $\xi_{\chi} (\xi^{'}_{\chi})$ and $\xi_{N} (\xi^{'}_{N})$ are two--component spinors of the DM fermion and nucleon,
$\hat{{\bf S}}_{\chi}$ and $\hat{{\bf s}}_{N}$ are spin operators of the LDCP and nucleon proportional to the Pauli matrices
whereas ${\bf q}$, ${\bf v}_{\chi}$ and ${\bf v}_{N}$ are three--dimensional vectors of momentum transfer, initial velocity of the DM particle
and initial velocity of the nucleon respectively. The first two terms in the right--hand side of Eq.~(\ref{18}), which are proportional
to $\mathcal{O}_1$ and $\mathcal{O}_5$, can be identified with spin--independent (SI) interactions of the Dirac DM fermions with nucleons. Two last terms
proportional to $\mathcal{O}_4$ and $\mathcal{O}_6$ are associated with the spin--dependent (SD) part of the $\chi$--nucleon scattering amplitude.
It is convenient to rewrite $\mathcal{M}_{\chi N}$ in the following form:
\begin{equation}
\mathcal{M}_{\chi N}\simeq 4 m_{\chi} m_N \left(\xi^{'\dagger}_{\chi} \xi^{'\dagger}_{N} (\hat{U}^N_{SI}+\hat{U}^N_{SD})\xi_{N} \xi_{\chi}\right)\,,
\label{191}
\end{equation}
\begin{equation}
\hat{U}^N_{SI}=\Biggl(\dfrac{e \mu_{\chi} F_e^N}{2 m_{\chi}} + \dfrac{m_N \varepsilon_H a_H^N}{m^2_H f}
- \dfrac{\bar{g} a^{\chi}_{V} a_V^N}{2 m^2_Z} \Biggr) \mathbb{I}_{\chi} \mathbb{I}_{N}
+ 2 i \dfrac{e \mu_{\chi} F_e^N}{|{\bf q}|^2} \left( \hat{{\bf S}}_{\chi} \cdot [{\bf q}\times {\bf v}_{\perp}]\right) \mathbb{I}_{N}\,,
\label{192}
\end{equation}
\begin{equation}
\hat{U}^N_{SD}=
\Biggl(\dfrac{2 \bar{g} a^{\chi}_{PV} a_{PV}^N}{m^2_Z}  + \dfrac{2 e \mu_{\chi} F_m^N}{m_N}\Biggr) (\hat{{\bf S}}_{\chi} \cdot \hat{{\bf s}}_{N})
- \dfrac{2 e \mu_{\chi} F_m^N}{|{\bf q}|^2 m_N} \left(\hat{{\bf S}}_{\chi} \cdot {\bf q}\right)\left(\hat{{\bf s}}_{N} \cdot {\bf q}\right)\,,
\label{193}
\end{equation}
where $\mathbb{I}_{\chi}$ and $\mathbb{I}_{N}$ are $2\times 2$ identity matrices.

Since the magnetic dipole moment $\mu_{\chi}$ of the Dirac DM fermions is strongly constrained \cite{PICO:2022ohk},
in our previous article we neglected $\mu_{\chi}$. When $\mu_{\chi}$ vanishes the SI part of $\mathcal{M}_{\chi N}$
is just a constant that does not depend on $|{\bf q}|^2$. As a consequence for $\mu_{\chi}\to 0$ the spin--independent LDCP--nucleon
scattering cross section $\sigma^N_{SI}$, i.e.
\begin{equation}
\sigma^N_{SI}\simeq\dfrac{m_{r}^2}{\pi} \Biggl|\dfrac{\varepsilon_H a_H^N m_N}{f m^2_{H}} -
\dfrac{\bar{g} a^{\chi}_{V} \langle a_{V} \rangle }{2 m_Z^2}\Biggr|^2 \,,
\qquad\qquad \langle a_{V} \rangle = \dfrac{1}{A}\Biggl( Z a^{p}_{V} + (A-Z) a^{n}_{V} \Biggr)\,,
\label{20}
\end{equation}
can be directly compared with the corresponding experimental limits, i.e. $\sigma^N_{SI\,\text{exp}}(m_{\chi})$ \cite{LZ:2024zvo}.
In Eq.~(\ref{20}) $A$ and $Z$ are the nucleon number and charge of the target nucleus.
The scenario under  consideration is not ruled out only if
\begin{equation}
\frac{\sigma^N_{SI\,\text{exp}}(m_{\chi})}{\sigma^N_{SI}}\gtrsim \xi=\frac{\rho_{\chi}}{\rho_{DM}}\,,
\label{21}
\end{equation}
where $\rho_{\chi}$ is the LDCP contribution to the total dark matter density $\rho_{DM}$.

\begin{figure}[htbp]
\centering
\begin{subfigure}[b]{0.47\textwidth}
\centering
\includegraphics[width=\textwidth]{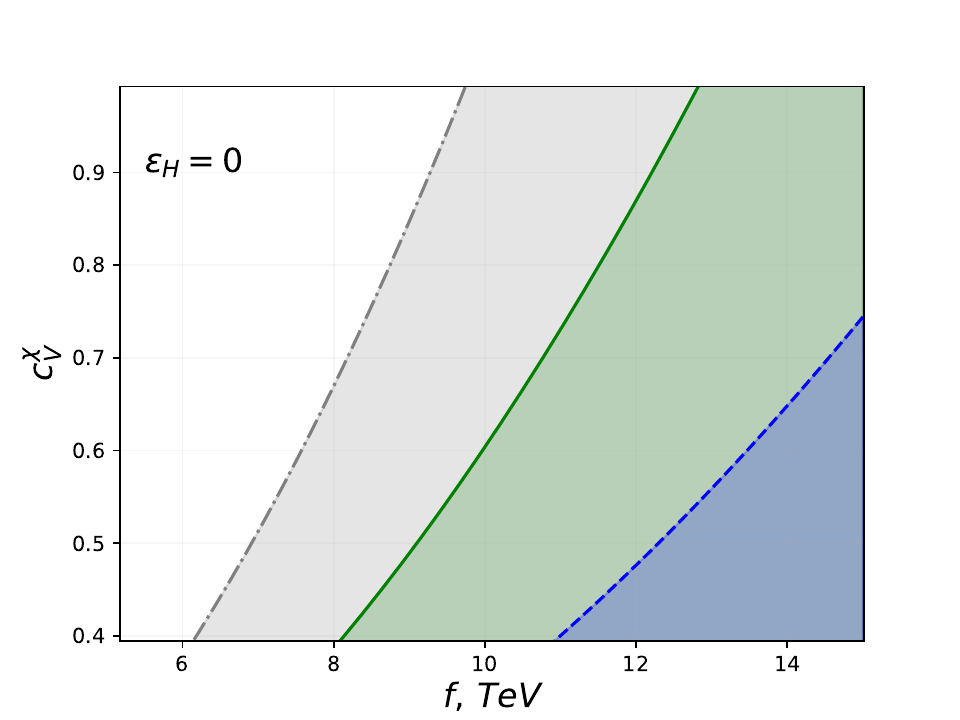}
\caption{}
\label{fig1:sub1}
\end{subfigure}
\hfill  
\begin{subfigure}[b]{0.47\textwidth}
\centering
\includegraphics[width=\textwidth]{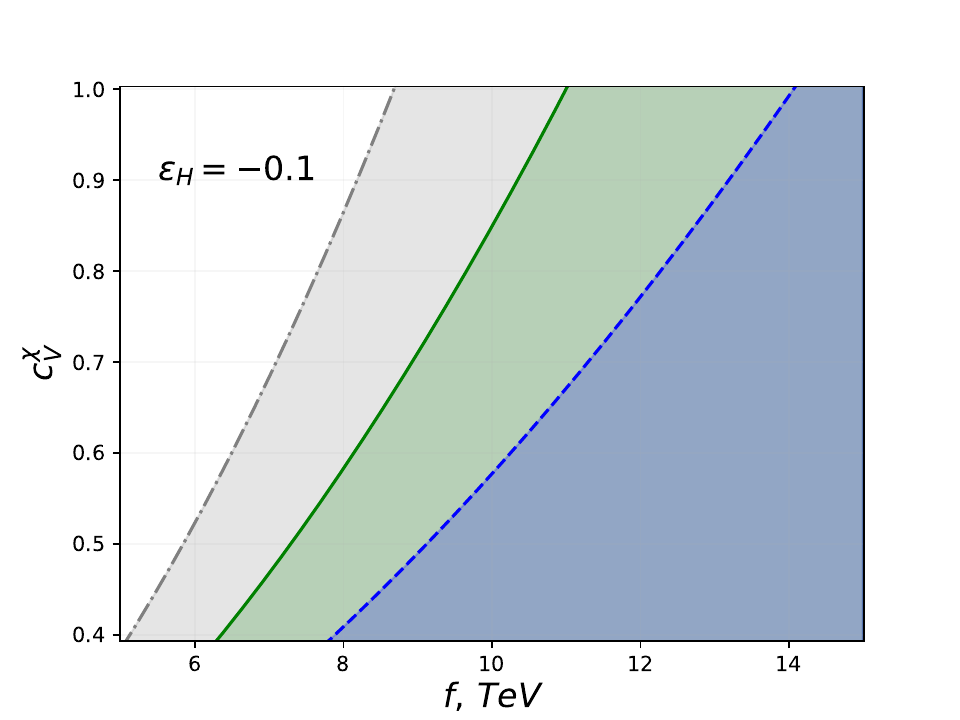}
\caption{}
\label{fig1:sub2}
\end{subfigure}
\vspace{0.2cm}
\begin{subfigure}[b]{0.47\textwidth}
\centering
\includegraphics[width=\textwidth]{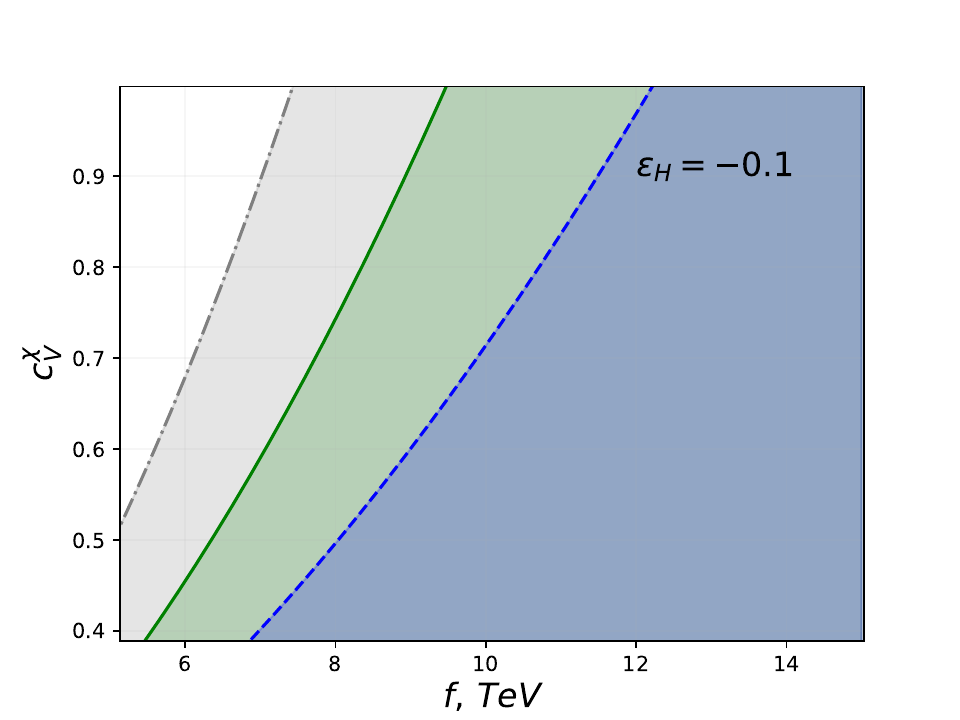}
\caption{}
\label{fig1:sub3}
\end{subfigure}
\hfill
\begin{subfigure}[b]{0.47\textwidth}
\centering
\includegraphics[width=\textwidth]{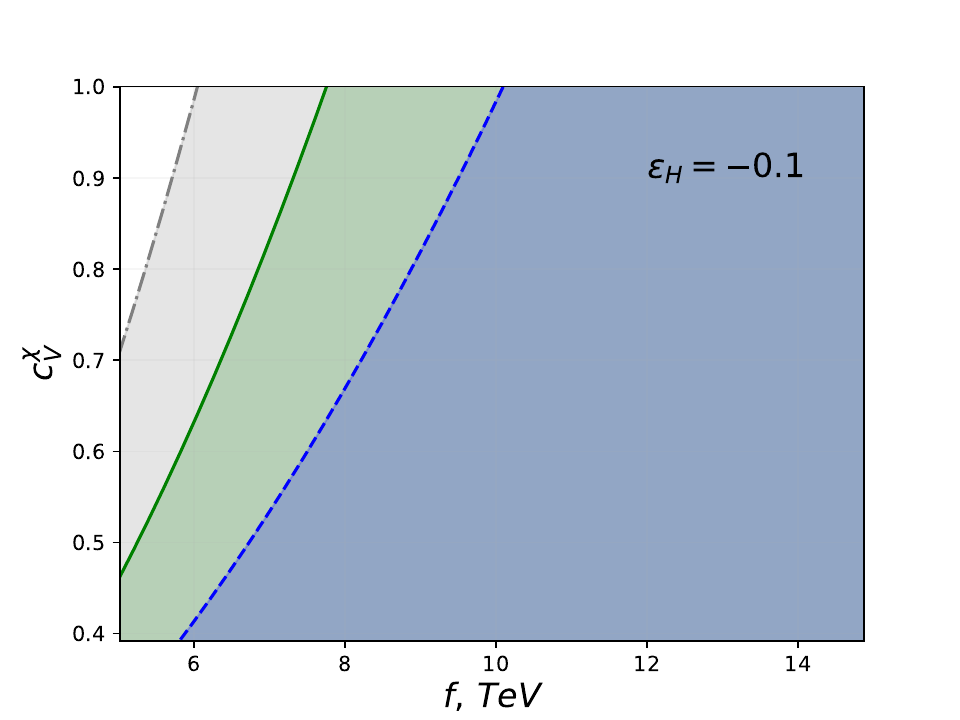}
\caption{}
\label{fig1:sub4}
\end{subfigure}
\caption{Different regions of the CHM parameter space in
the $f - c^{\chi}_{V}$ plane for $m_{\chi}=200\,\mbox{GeV}$ and
$\varepsilon_{H}=0$ {\it (a)}, for $m_{\chi}=200\,\mbox{GeV}$ and
$\varepsilon_{H}=-0.1$ {\it (b)}, for $m_{\chi}=500\,\mbox{GeV}$ and
$\varepsilon_{H}=-0.1$ {\it (c)} as well as for $m_{\chi}=1\,\mbox{TeV}$ and
$\varepsilon_{H}=-0.1$ {\it (d)}. Below dashed--dotted, solid and dashed lines
$\rho_{\chi}$ is allowed to be bigger than $0.1\cdot \rho_{DM}$,
can be larger than $0.3\cdot \rho_{DM}$ and may constitute full dark matter
density $\rho_{DM}$ respectively.}
\label{fig1.1}
\end{figure}

In contrast to our previous paper \cite{Belyakova:2024fcw} here we use condition (\ref{21}) to constrain the value of $\rho_{\chi}$.
We restrict our consideration to the scenarios with $\xi\ge 0.1$. For $f\lesssim 15\,\mbox{TeV}$ and $|\varepsilon_H|\le 0.1$
the interactions of the LDCP with nucleons are dominated by the $t$-channel exchange of the $Z$ boson.
Because the corresponding contribution to the spin--independent part of $\mathcal{M}_{\chi N}$ is determined by
$c_V^{\chi}$ and $f$ (see Eqs.~(\ref{8}) and (\ref{18})) in Fig.~1 we identified the regions of the parameter space
in the $c_V^{\chi} - f$ plane where $\xi=1$ ($\rho_{\chi}$ = $\rho_{DM}$), as well as where $\xi$ is allowed to be larger than $0.1$ and $0.3$.
The compositeness scale $f$ is varied from $5\,\mbox{TeV}$ to $15\,\mbox{TeV}$ while $c_V^{\chi}$ changes from $0.5$ to $1$.

From Figs.~1a and 1b it follows that the constraint on the parameter space caused by the requirement $\xi\ge 0.1$ becomes
weaker for negative values of $\varepsilon_H$. Nevertheless for $f\simeq 5\,\mbox{TeV}$ and $m_{\chi} < 500\,\mbox{GeV}$
the value of $\xi$ cannot exceed $0.1$. Since $\sigma^N_{SI}$ diminishes with increasing $f$ the allowed range of $c_V^{\chi}$
enlarges with the growth of the compositeness scale. Moreover Figs. 1b, 1c and 1d demonstrate that near $f\simeq 10\,\mbox{TeV}$
and $c_V^{\chi}\simeq 0.5$ for $m_{\chi} > 200\,\mbox{GeV}$ there is always some part of the parameter space where $\rho_{\chi}$
can constitute full dark matter density. Thus for $f\simeq 5\,\mbox{TeV}$ we further consider $m_{\chi}\gtrsim\, 500\,\mbox{GeV}$
whereas for $f\simeq 10\,\mbox{TeV}$ and $f\simeq 15\,\mbox{TeV}$ we explore scenarios with $m_{\chi}\gtrsim\,200\,\mbox{GeV}$.

\begin{table}[htbp]
\centering
\caption{Parameters and $\sigma^N_{SI}$ for different benchmark scenarios as well
as the corresponding experimental bounds $\sigma^N_{SI\,\text{exp}}$ and $\mu^{exp}_{\chi}$}.
\label{tab:BM}
\begin{tabular}{c c c c c c c}
\toprule
$m_{\chi}$ [GeV]  &  $f$ [TeV] & $c_V^{\chi}$ & $\varepsilon_H$ & $\sigma^N_{SI}$ [yb]& $\sigma^N_{SI\,\text{exp}}$ [yb]
& $\mu^{exp}_{\chi}$ [$\text{GeV}^{-1}$]\\
\midrule
      & $5$  &  $0.5$ & $-0.1$    & $117.2$    &     &\\
      & $10$ &  $0.5$ & $0$       & $12.9$     &     & \\
$500$ & $10$ &  $0.5$ & $-0.1$    & $3.3$      &  12 & $10^{-8}$\\
      & $15$ &  $0.5$ & $0$       & $2.43$     &     & \\
      & $15$ &  $0.5$ & $-0.1$    & $0.16$     &     & \\
\midrule
       & $5$  &  $0.5$ & $-0.1$    & $117.4$   &     & \\
       & $10$ &  $0.5$ & $0$       & $13$      &     & \\
$1000$ & $10$ &  $0.5$ & $-0.1$    & $3.3$     &  30 & $1.5\cdot 10^{-8}$\\
       & $15$ &  $0.5$ & $0$       & $2.43$    &     & \\
       & $15$ &  $0.5$ & $-0.1$    & $0.16 $   &     & \\
\midrule
       & $10$ &  $0.5$ & $0$       & $12.9$    &     & \\
$200$  & $10$ &  $0.5$ & $-0.1$    & $3.3$     &   6 & $0.8\cdot 10^{-8}$\\
       & $15$ &  $0.5$ & $0$       & $2.43$    &     & \\
       & $15$ &  $0.5$ & $-0.1$    & $0.16 $   &     & \\
\bottomrule
\end{tabular}
\end{table}

In Table~\ref{tab:BM} a set of benchmark scenarios is specified. All these scenarios remain phenomenologically viable
when $\xi=0.1$\,. The current experimental limits on the magnetic moment of the DM fermions $\mu^{exp}_{\chi}$ and
$\sigma^N_{SI\,\text{exp}}$ are given in the last two columns of this Table. In the next Section the benchmark scenarios
presented in Table~\ref{tab:BM} are going to be used to illustrate the results of our numerical analysis of the differential
cross section of the LDCP--nucleus scattering and the corresponding event rate.

Concerning the SD LDCP-nucleon interaction, the averaged value of the corresponding amplitude squared also does not depend on $|{\bf q}|^2$
in the non-relativistic limit. Therefore, the SD scattering cross section
\begin{equation}
\sigma^N_{SD}\simeq\dfrac{m_{r}^2}{2\pi} \Biggl[ \dfrac{3}{2} \left(\dfrac{\bar{g} a^{\chi}_{PV} a^N_{PV}}{m_Z^2} \right)^2
+2\left(\dfrac{\bar{g} a^{\chi}_{PV} a^N_{PV}}{m_Z^2} \right) \left(\dfrac{e\mu_{\chi}}{m_N} F_m^N\right)
+ \left(\dfrac{e\mu_{\chi}}{m_N} F_m^N\right)^2
\Biggr] \,,
\label{22}
\end{equation}
can be directly compared with the experimental bounds on LDCP-proton and LDCP-neutron scattering cross
sections, i.e. $\sigma^p_{exp}$ and $\sigma^n_{exp}$. The computed values of $\sigma^N_{SD}$ (\ref{22}) are determined by three parameters:
$f$,~ $c^{\chi}_{PV}\sim 1$~ and~ $\mu_{\chi}$. For $m_{\chi}\simeq 200 - 1000\,\text{GeV}$ the reduced mass $m_r \approx m_N$.
Thus $\sigma^N_{SD}$ does not depend significantly on the LDCP mass.

\begin{figure}[htbp]
\centering
\begin{subfigure}[b]{0.47\textwidth}
\centering
\includegraphics[width=\textwidth]{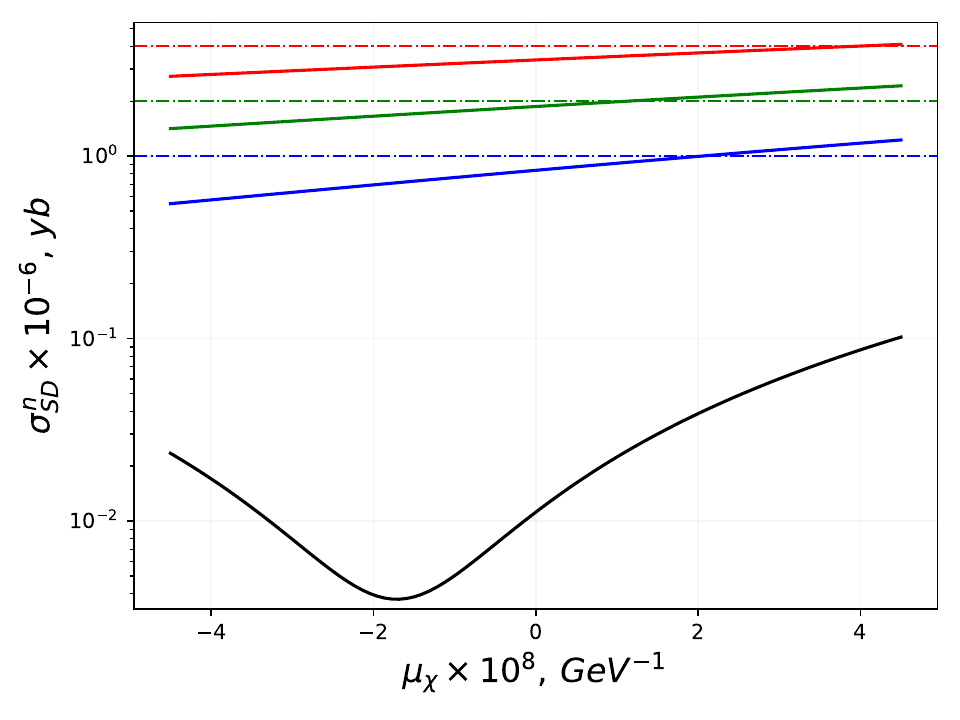}
\caption{}
\label{fig2:sub1}
\end{subfigure}
\hfill  
\vspace{0.2cm}
\begin{subfigure}[b]{0.47\textwidth}
\centering
\includegraphics[width=\textwidth]{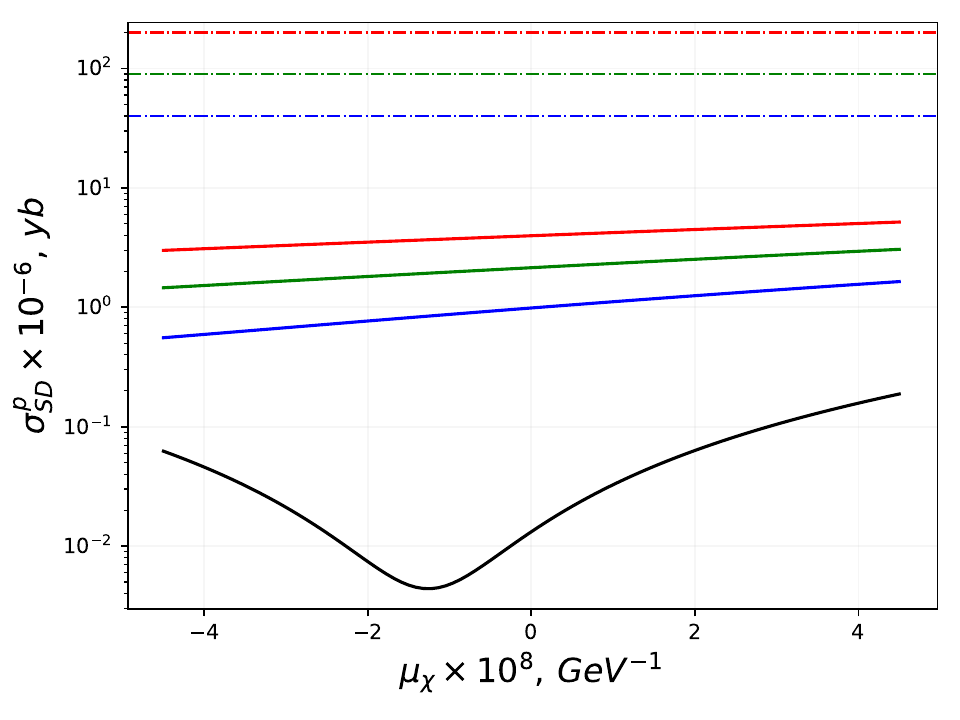}
\caption{}
\label{fig2:sub3}
\end{subfigure}
\hfill
\caption{Spin--dependent LDCP--neutron {\it (a)} and LDCP--proton {\it (b)} scattering cross sections as a function of the
LDCP magnetic dipole moment $\mu_{\chi}$ for $c^{\chi}_{PV}=1$. Solid black, blue, green and red lines correspond to
$f=5\,\mbox{TeV}$, $f=1.7\,\mbox{TeV}$, $f=1.4\,\mbox{TeV}$ and $f=1.2\,\mbox{TeV}$ respectively. The horizontal dashed--dotted
red, green and blue lines represent the experimental limits on the appropriate SD cross sections for $m_{\chi} = 1000\,\mbox{GeV}$,
$m_{\chi} = 500\,\mbox{GeV}$ and $m_{\chi} = 200\,\mbox{GeV}$.}
	\label{fig2.1}
\end{figure}

In Figs.~\ref{fig2:sub1} and \ref{fig2:sub3} the LDCP-neutron and LDCP-proton cross sections ($\sigma^n_{SD}$ and $\sigma^p_{SD}$)
are plotted as a function of the LDCP magnetic dipole moment $\mu_{\chi}$ for $c^{\chi}_{PV}=1$ and different values of the compositeness scale $f$.
Solid black, blue, green and red lines represent $\sigma^n_{SD}$ and $\sigma^p_{SD}$ calculated for $f=5\,\mbox{TeV}$,
$f=1.7\,\mbox{TeV}$, $f=1.4\,\mbox{TeV}$ and $f=1.2\,\mbox{TeV}$ respectively. The computed the LDCP-neutron and LDCP-proton
cross sections are compared with the corresponding experimental bounds for $m_{\chi} = 1000\,\mbox{GeV}$,
$m_{\chi} = 500\,\mbox{GeV}$ and $m_{\chi} = 200\,\mbox{GeV}$. As follows from Figs.~\ref{fig2:sub1} and \ref{fig2:sub3}
for $f\gtrsim 5\,\mbox{TeV}$ the SD LDCP-nucleon cross sections are considerably lower than the current experimental limits. Nevertheless
$\sigma^p_{exp}$ and $\sigma^n_{exp}$ may provide an important lower bound on the compositeness scale. Indeed, Eq.~(\ref{20}) indicates
that the SM--like Higgs and $Z$ boson contributions to the SI LDCP--nucleon cross section can cancel each other. Therefore even
for $f\lesssim 5\,\mbox{TeV}$ the value of $\sigma^N_{SI}$ can be significantly smaller than $\sigma^N_{SI\,\text{exp}}(m_{\chi})$.
In this case the lower bound on the compositeness scale is set by $\sigma^n_{exp}(m_{\chi})$ because
$\sigma^n_{exp}(m_{\chi})\ll \sigma^p_{exp}(m_{\chi})$. Fig.~\ref{fig2:sub1} demonstrates that for $\xi\simeq 1$ the lower bound
on $f$ varies from $f=1.7\,\mbox{TeV}$ to $f=1.2\,\mbox{TeV}$ when $m_{\chi}$ changes from $m_{\chi} = 200\,\mbox{GeV}$ to
$m_{\chi} = 1\,\mbox{TeV}$.

\section{Interaction of Dirac DM fermions with Xe and Ar nuclei}

\subsection{Differential cross section}

Using the requirement $\xi\ge 0.1$, as well as the constraints on the parameter space that come from the experimental limits
on $\sigma^N_{SI}$ and $\mu_{\chi}$, we proceed to investigate LDCP scattering on Xenon (LZ, XENONnT and PandaX experiments)
and Argon (DarkSide Project) nuclei. Characteristics of the most abundant Xenon and Argon isotopes are summarized
in Table~\ref{tab:xe_isotopes} (see also \cite{Banks:2010eh}).

\begin{table}[htbp]
\centering
\caption{Most common isotopes of Xe and Ar, their abundances $f_A$, spin $J$, magnetic moments $\mu_T^A$
in units of nuclear magneton $\mu_n=e/2m_N$ as well as intervals of $\mathcal{S}_p$ and $\mathcal{S}_n$\,.}
\label{tab:xe_isotopes}
\begin{tabular}{l c c c c c c}
\toprule
Isotope & Z & Spin $J$ & abundance $f_A$ & $\mu_T^A/\mu_n$ & $\mathcal{S}_p$ & $\mathcal{S}_n$\\
\midrule
$\ce{^{132}Xe}$ & 54 & $0$ & $26.9\%$ & - & - & - \\
$\ce{^{129}Xe}$ & 54 & $1/2$ & $26.4$\% & -0.778 & [-0.002, 0.028] & [0.248, 0.359]\\
$\ce{^{131}Xe}$ & 54 & $3/2$ & $21.2\%$ & +0.692 & [-0.012, -0.0007] & [-0.272, -0.125] \\
$\ce{^{134}Xe}$ & 54 & $0$ & $10.4\%$ & - & - & -  \\
$\ce{^{136}Xe}$ & 54 & $0$ & $8.9\%$ & - & - & - \\
$\ce{^{130}Xe}$ & 54 & $0$ & $4.1\%$ & - & - & - \\
$\ce{^{128}Xe}$ & 54 & $0$ & $1.9\%$ & - & - & - \\
$\ce{^{40}Ar}$ & 18 & $0$ & $99.6\%$ & - & - & - \\
$\ce{^{36}Ar}$ & 18 & $0$ & $0.3\%$ & - & - & - \\		
\bottomrule
\end{tabular}
\end{table}

In the non-relativistic limit the differential cross section of the LDCP--nucleus scattering can be
presented in the following form
\begin{equation}
\dfrac{d\sigma}{dy}=\dfrac{d\sigma_{SI}}{dy}+\dfrac{d\sigma_{SD}}{dy}=
\frac{M_{r}^2}{\pi}\Biggl(\langle \hat{U}_{SI} \hat{U}_{SI}^+\rangle + \langle \hat{U}_{SD} \hat{U}_{SD}^+\rangle\Biggr),
\label{23}
\end{equation}
$$
y=\frac{\varepsilon_{rec}}{\varepsilon_{max}},\qquad M_r=\frac{m_T m_{\chi}}{m_T + m_{\chi}},
$$
where
$m_T$ is a nucleus mass, $\varepsilon_{rec}$ and $\varepsilon_{max}$ are the recoil energy of nuclei and
its maximal value in the laboratory frame, i.e. $\varepsilon_{rec}=|{\bf q}|^2/2 m_T$ and $\varepsilon_{max}=2 M_r^2 |{\bf v}_{\chi}|^2/m_T$,
while $|{\bf v}_{\chi}|\sim 10^{-3}$ is the LDCP speed in the laboratory frame.
The SI part of the LDCP--nucleus interaction potential $\hat{U}_{SI}$ is given by
\begin{eqnarray}
\hat{U}_{SI}({\bf q}^2)\simeq \left[\left(\frac{Ze\mu_{\chi}}{2m_{\chi}}
-\frac{A\bar{g}\langle a_V\rangle a_V^{\chi}}{2m_Z^2}+\frac{A a_H^N \varepsilon_H m_N}{m_H^2 f}\right)
\mathbb{I}_{\chi} \mathbb{I}_{T}\qquad\qquad\qquad\right. \nonumber \\
\left.\qquad\qquad\qquad+\mu_{\chi}\hat{{\bf S}}_{\chi}\cdot \left[i{\bf q}\times {\bf v}_{\chi}\right]\frac{2Ze}{|{\bf q}|^2}\mathbb{I}_{T}
\right]F_{SI}(|{\bf q}|^2)\,.
\label{eq3.100}
\end{eqnarray}
For $|{\bf q}|\to 0$ the value of $\hat{U}_{SI}=\sum_{N} \hat{U}^N_{SI}$, where one needs to sum over all nucleons $N$ comprising the nucleus
replacing $\mathbb{I}_{N}$ by $\mathbb{I}_{T}$. In Eq.~(\ref{eq3.100}) $\mathbb{I}_{T}$ is an identity matrix, $Z$ and $A$ are nucleus charge
and mass number respectively. Nuclear SI form factor $F_{SI}(|{\bf q}|^2)$ is normally parametrized as
\begin{equation}
F_{SI}({\bf q}^2)=3\left[\frac{\sin(|{\bf q}|r)- (|{\bf q}|r) \cos(|{\bf q}|r)}{\left(|{\bf q}|r\right)^3}\right]\exp^{-|{\bf q}|^2 s^2},
\label{f_SI}
\end{equation}
where $r=1.12 A^{1/3}$ fm and $s$ = 1 fm \cite{Hambye:2021xvd}. Substituting (\ref{eq3.100}) into Eq.~(\ref{23}) one finds
\begin{equation}
\frac{d\sigma_{SI}}{dy}\simeq	\frac{{M_{r}}^2 F^2_{SI}(|{\bf q}|^2)}{\pi}
\left[\left(\frac{Ze\mu_{\chi}}{2m_{\chi}}-\frac{A\bar{g}\langle a_V\rangle a_V^{\chi}}{2 m_Z^2}+
\frac{A\varepsilon_H a_H^N m_N}{f m_H^2}\right)^2+\frac{\left(e\mu_{\chi}Z\right)^2}{4M_r^2}\left(\frac{1}{y}-1\right)\right]\,.
\label{eq3.12}
\end{equation}

The SD part of the LDCP--nucleus interaction potential $\hat{U}_{SD}$ can be written as
\begin{equation}
\hat{U}_{SD}({\bf q}^2)\simeq \biggl\{\left(\hat{{\bf S}}_{\chi}\hat{{\bf J}}\right)\frac{c_4}{J} - \left(\hat{{\bf S}}_{\chi} {\bf q}\right)
\left(\hat{{\bf J}}{\bf q}\right)\dfrac{e\mu_{\chi}\tilde{\mu}_T}{{\bf q}^2m_N J}\biggr\}F_{SD}({\bf q}^2)\,,
\label{27}
\end{equation}
\begin{equation}
c_4=\mathcal{S}_p \left[\dfrac{2\bar{g}a_{PV}^pa_{PV}^{\chi}}{m_Z^2}\right]+
\mathcal{S}_n \left[\dfrac{2\bar{g}a_{PV}^na_{PV}^{\chi}}{m_Z^2}\right]
+\dfrac{e\mu_{\chi}\tilde{\mu}_T}{m_N}\,,
\label{28}
\end{equation}
\begin{equation}
\hat{{\bf J}}=\hat{{\bf J}}_p + \hat{{\bf J}}_n + \hat{{\bf L}}_p + \hat{{\bf L}}_n\,,
\qquad\qquad \hat{{\bf J}}_p =\sum_{i} \hat{{\bf s}}_{pi}\,,\qquad\qquad
\hat{{\bf J}}_n =\sum_{i} \hat{{\bf s}}_{ni}\,,
\label{29}
\end{equation}
where the index $i$ runs over all protons (neutrons) in the nucleus, $\hat{{\bf J}}$ is an operator of total angular momentum of nucleus,
$\hat{{\bf J}}_p$ and $\hat{{\bf J}}_n$ are operators of the total spin of protons and neutrons, while
$\hat{{\bf L}}_p$ and $\hat{{\bf L}}_n$ are operators of the total orbital momentum of protons and neutrons forming nucleus.
In the derivation process, it is usually assumed that
$$
\langle T'|\hat{\bf J}_p|T\rangle=\frac{\mathcal{S}_p}{J}\langle T'|\hat{{\bf J}}|T\rangle\,, \qquad
\langle T'|\hat{\bf J}_n|T\rangle=\frac{\mathcal{S}_n}{J}\langle T'|\hat{{\bf J}}|T\rangle\,, \qquad
\langle T'|\hat{\bf L}_{p,n}|T\rangle=\frac{\mathcal{L}_{p,n}}{J}\langle T'|\hat{{\bf J}}|T\rangle\,.
$$
The intervals of variations of $\mathcal{S}_p$ and $\mathcal{S}_n$ are specified in Table~\ref{tab:xe_isotopes} (see also \cite{DelNobile:2021wmp}).
In Eqs.~(\ref{27})--(\ref{28}) $\tilde{\mu}_T$, which is the value of the magnetic moment of nucleus in units of nuclear magneton,
is defined as
$$
\tilde{\mu}_T= 2 F_m^p \mathcal{S}_p + 2 F_m^n \mathcal{S}_n + \mathcal{L}_p\,.
$$
Nuclear SD form factor $F_{SD}(|{\bf q}|^2)$ in Eq.~(\ref{27}) is often approximated as
\begin{equation}
F_{SD}({\bf q}^2)=\frac{\sin(|{\bf q}|\tilde{r})}{\left(|{\bf q}|\tilde{r}\right)}\Theta(|{\bf q}|\tilde{r}<2.55)
+0.21\Theta(2.55<|{\bf q}|\tilde{r}<4.5),
\label{30}	
\end{equation}
where $\tilde{r}= A^{1/3}$ fm \cite{Hambye:2021xvd}. In the limit $|{\bf q}|\to 0$
the expressions (\ref{27})--(\ref{29}) can be obtained by summing all SD $\chi$--nucleon interactions, i.e.
$\hat{U}_{SD}=\sum_N \hat{U}^N_{SD}$.

The SD part of the differential cross section of the LDCP--nucleus scattering takes the form:
\begin{equation}
\frac{d\sigma_{SD}}{dy}\simeq\frac{M_r^2 F_{SD}^2({\bf q}^2)}{\pi}\frac{(J+1)}{12 J}
\left[\left(c_4-\dfrac{e\mu_{\chi}\tilde{\mu}_T}{m_N}\right)^2 + 2 c_4^2\right]\,.
\label{SDcross}
\end{equation}
In general this part of the differential cross section depends on $\mathcal{S}_p$, $\mathcal{S}_n$ and $\mathcal{L}_p$,
which vary from one nuclear model to another. In particular, from Table~\ref{tab:xe_isotopes} it follows that for
the Xenon nuclei $|\mathcal{S}_p|\ll |\mathcal{S}_n|$. At the same time the results of our numerical analysis indicate that
in the case of the LDCP--Xe scattering the contribution of the SD part to the total differential cross section tends to be much
smaller than the contribution of the SI one. Consequently, the choice of a particular nuclear model does not significantly affect the
corresponding scattering cross section. For the subsequent analysis of the LDCP scattering on Xe isotopes with non-zero
spin, we use the Odd Group Model (OGM) \cite{Goodman:1984dc,Engel:1989ix} (for a review, see \cite{Bednyakov:2004xq}),
which implies that only one unpaired nucleon contribute to the nucleus spin $J$.
Thus for Xe nuclei, which possess one unpaired neutron, we set
$$
\mathcal{S}_p = 0\,,\qquad\qquad \mathcal{L}_p = 0\,,\qquad\qquad \mathcal{S}_n = \dfrac{\tilde{\mu}_T}{2F_m^n}\,.
$$
Since the most abundant Argon isotopes have $J=0$ (see Table~\ref{tab:xe_isotopes}) the SD part of the differential cross section
of the LDCP--Ar scattering vanishes.
Although in this paper a simplified approach for the calculation of the differential cross section is adopted
we compare it with the results obtained by using response-function formalism \cite{Fitzpatrick:2012ix}
(see Tabels \ref{tab:dsigma} and \ref{tab:fractions}).
As follows from Table \ref{tab:dsigma} at low recoil energies of nuclei these two approaches lead to almost identical results
for the SD part of the differential cross section of the LDCP--Xe scattering.

\begin{table}[htbp]
\centering
\caption{Spin-independent and spin-dependent contributions to the differential cross section of LDCP--Xe scattering
at \(\varepsilon_{\rm rec}=2\) KeV for $|{\bf v}_{\chi}|= 10^{-3}$, $c^{\chi}_V=0.5$, \(c_{PV}^\chi=0.5\) and \(\mu_\chi=3\,\mu_\chi^{\rm exp}(m_\chi)\).
\(d\sigma_{SD}/dy\) is computed with the simplified SD form factor, Eq.~(\ref{30}), while for the calculation of \(d\sigma^{\rm NR}_{SD}/dy\)
the nuclear response functions of Ref.~\cite{Fitzpatrick:2012ix} are used.}
\label{tab:dsigma}
\begin{tabular}{cccccc}
\toprule
\(m_\chi\) [GeV] & \(f\) [TeV] & \(\varepsilon_H\) & \(d\sigma_{SI}/dy\) [fb] & \(d\sigma_{SD}/dy\) [fb] & \(d\sigma^{\rm NR}_{SD}/dy\) [fb] \\
\midrule
     & 10 & 0 & 1.6 & \(4.9\cdot 10^{-3}\)   & \(4.8\cdot 10^{-3}\) \\
200	 & 10 & -0.1 & 0.55 &                      &                      \\
	 & 15 & 0    & 0.48 & \(4.4\cdot 10^{-3}\)   & \(4.3\cdot 10^{-3}\) \\
	 & 15 & -0.1 & 0.21 &                      &                      \\
\midrule
     & 5  & -0.1 & 20.2 & \(1.8\cdot 10^{-2}\) & \(1.9\cdot 10^{-2}\) \\
     & 10 & 0 & 2.8 & \(1.2\cdot 10^{-2}\) & \(1.2\cdot 10^{-2}\) \\
500	 & 10 & -0.1 & 1.1 &                      &                      \\
	 & 15 & 0    & 0.97 & \(1.1\cdot 10^{-2}\) & \(1.1\cdot 10^{-2}\) \\
	 & 15 & -0.1 & 0.54 &                      &                      \\
\midrule
     &  5 & -0.1 & 25.6 & \(4.2\cdot 10^{-2}\) & \(4.3\cdot 10^{-2}\) \\
     & 10 & 0    & 4.2 & \(3.2\cdot 10^{-2}\) & \(3.15\cdot 10^{-2}\) \\
1000 & 10 & -0.1 & 2.13 &                      &                      \\
	 & 15 & 0    & 2.0  & \(3.0\cdot 10^{-2}\) & \(2.96\cdot 10^{-2}\) \\
	 & 15 & -0.1 & 1.5  &                      &                      \\
\bottomrule
\end{tabular}
\end{table}

Here we analyse the differential cross section of the LDCP--nucleus scattering averaged over isotopes, i.e.
\begin{equation}
\dfrac{d\sigma}{dy}=\sum_{A} f_A \dfrac{d\sigma^A}{dy}=\sum_{A} f_A \dfrac{d\sigma^A_{SI}}{dy} +
\sum_B f_B \dfrac{d\sigma^B_{SD}}{dy}\,,
\label{32}
\end{equation}
where index $A$ runs over all isotopes and index $B$ corresponds to the isotopes with non--zero spin.
The abundances $f_A$ of the most common isotopes of Xe and Ar are specified in Table~\ref{tab:xe_isotopes}.
Because $|{\bf q}|^2=4 M^2_r |{\bf v}_{\chi}|^2 y$, the differential cross section (\ref{32}) depends on
eight variables: $m_{\chi}$, $f$, $c^{\chi}_V$, $c^{\chi}_{PV}$, $\varepsilon_{H}$, $\mu_{\chi}$, $|{\bf v}_{\chi}|$ and $y$.	
The smallness of the SD part of the differential cross section implies that $\dfrac{d\sigma}{dy}$ does not change much
when $c^{\chi}_{PV}$ is varied. To simplify our analysis we fix
$c^{\chi}_{PV}=0.5$ and $|{\bf v}_{\chi}|=10^{-3}$.
Then the differential cross section (\ref{32}) remains a function of
\begin{equation}	
m_{\chi}\,,\qquad f\,,\qquad c^{\chi}_V\,,\qquad \varepsilon_{H}\,,\qquad \mu_{\chi}\,, \qquad y\quad(\mbox{or}\quad \varepsilon_{rec})\,.
\label{33}
\end{equation}

\begin{figure}[htbp]
\centering
\begin{subfigure}[b]{0.47\textwidth}
\hspace{-3.8cm}\centering
\includegraphics[width=1\textwidth]{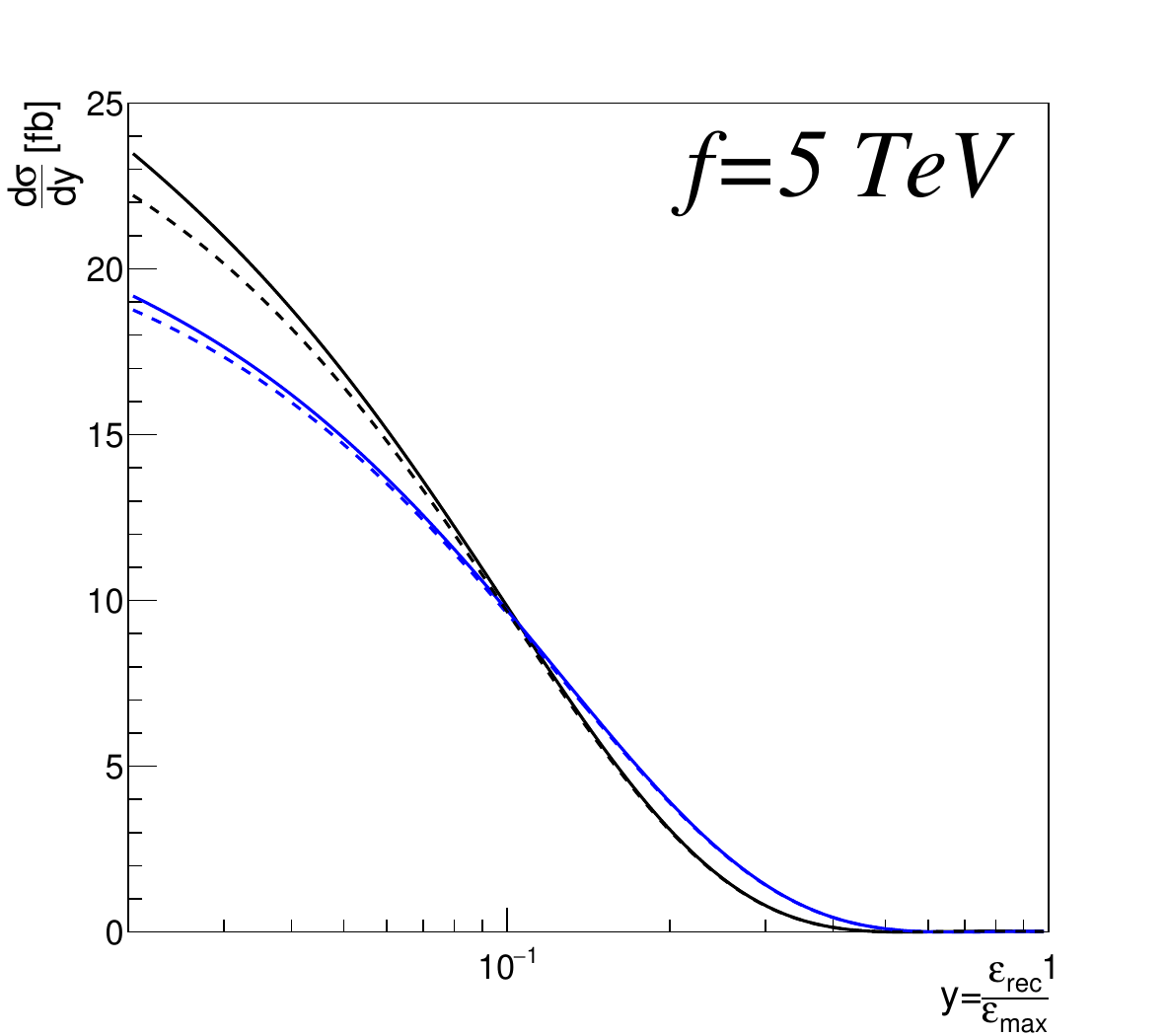}
\caption{}
\label{fig1:sigma_sub1}
\end{subfigure}
\hspace{-1cm}
\begin{subfigure}[b]{0.47\textwidth}
\centering
\includegraphics[width=1\textwidth]{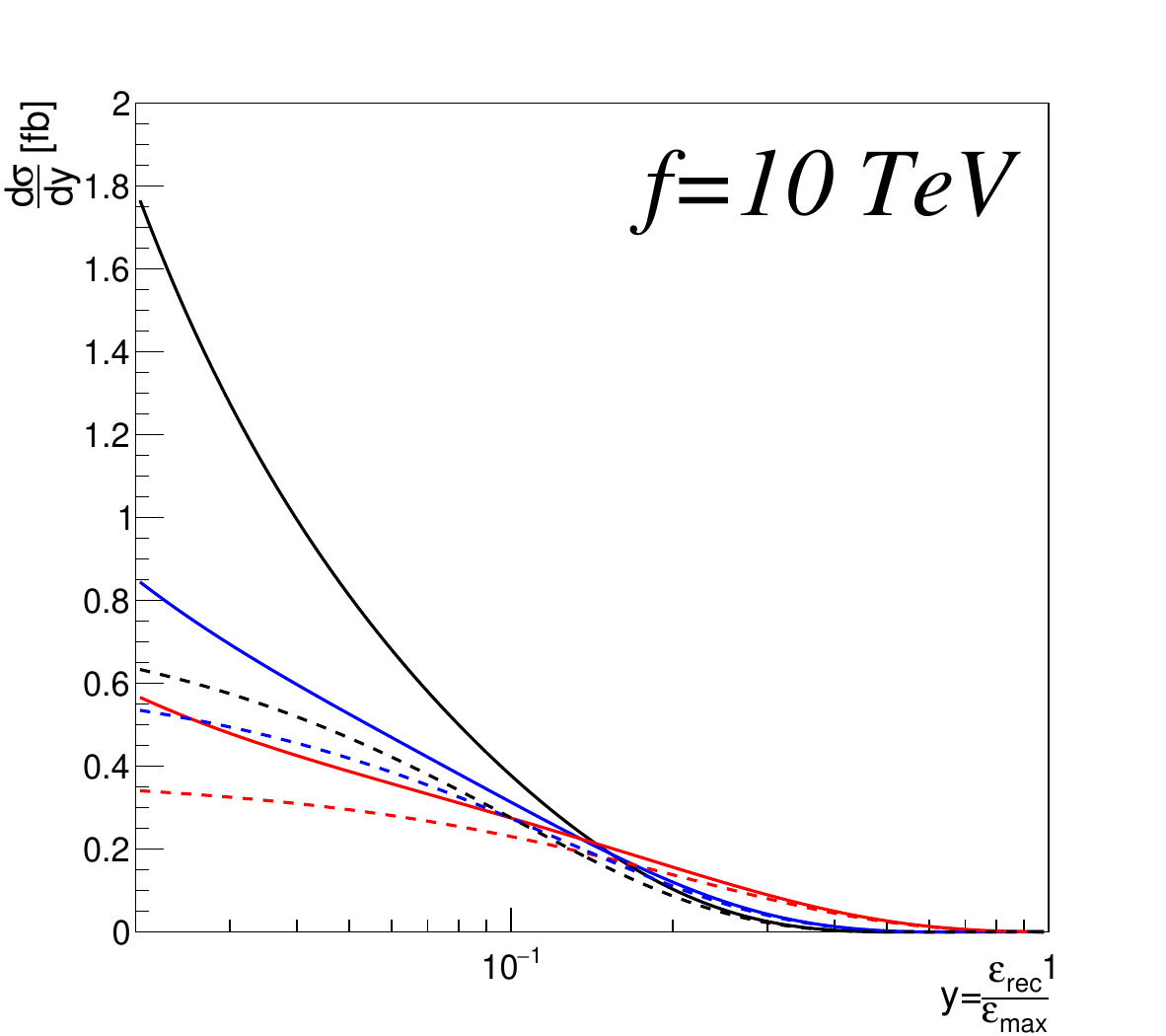}
\caption{}
\label{fig1:sigma_sub2}
\end{subfigure}
\caption{Differential cross section of the LDCP-Xe scattering averaged over Xe isotopes as a function of $y$ for $f=5\,\mbox{TeV}$, $c_V^{\chi}=0.5$
and $\varepsilon_H=-0.1$ {\it (a)} as well as for $f=10\,\mbox{TeV}$, $c_V^{\chi}=0.5$ and $\varepsilon_H=-0.1$ {\it (b)}.
Red, blue and black lines are associated with $m_{\chi}=200\,\mbox{GeV}$, $m_{\chi}=500\,\mbox{GeV}$ and $m_{\chi}=1000\,\mbox{GeV}$, respectively.
Dashed and solid lines correspond to $\mu_{\chi}=0$ and $\mu_{\chi}= 3\cdot \mu_{\chi}^{\text{exp}}(m_{\chi})$.}
\label{fig2.0}
\end{figure}

\begin{figure}[htbp]
\centering
\begin{subfigure}[b]{0.47\textwidth}
\hspace{-3.8cm}\centering
\includegraphics[width=1\textwidth]{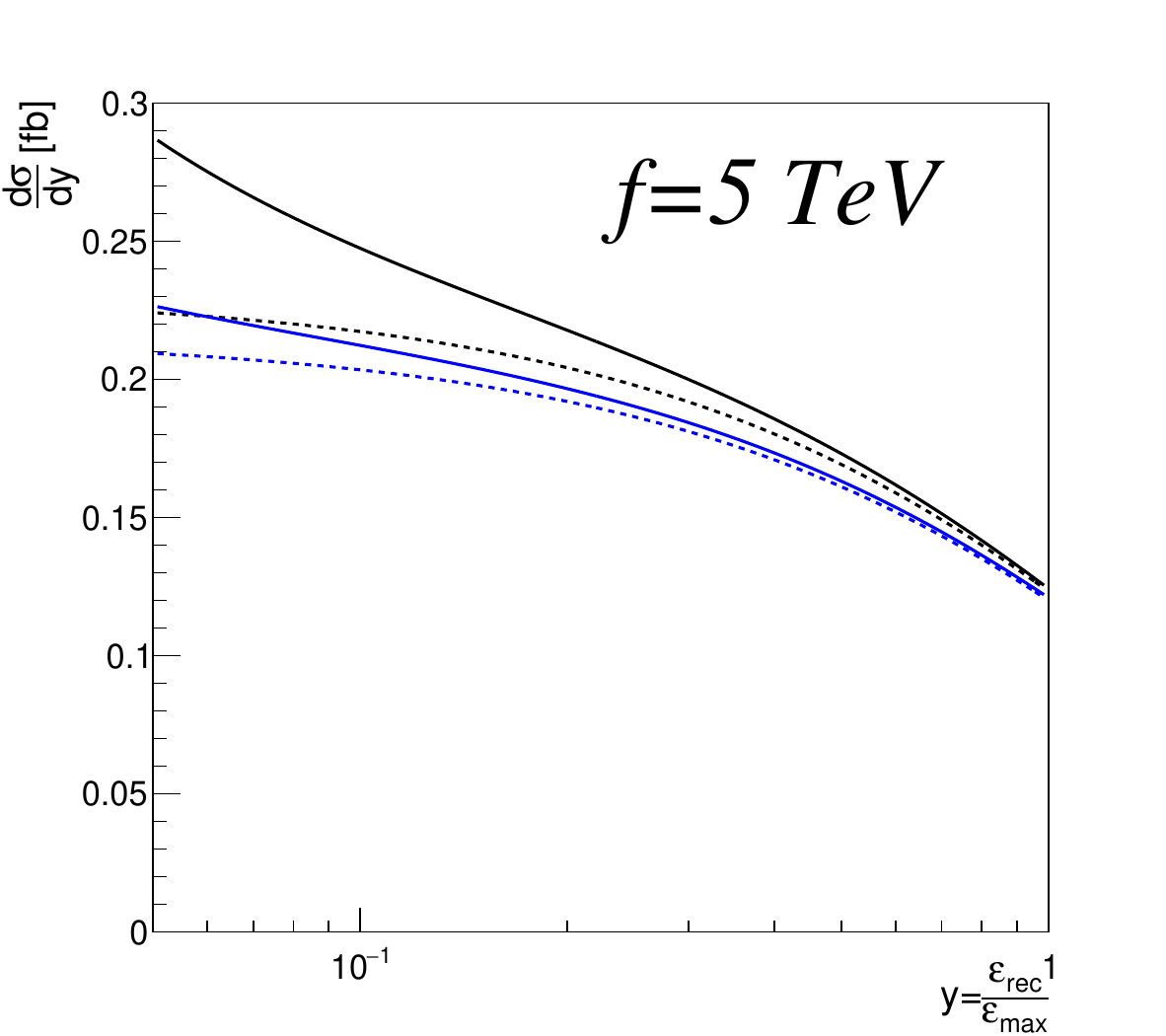}
\caption{}
\label{fig1:sigmaAr_sub1}
\end{subfigure}
\hspace{-1cm}
\begin{subfigure}[b]{0.47\textwidth}
\centering
\includegraphics[width=1\textwidth]{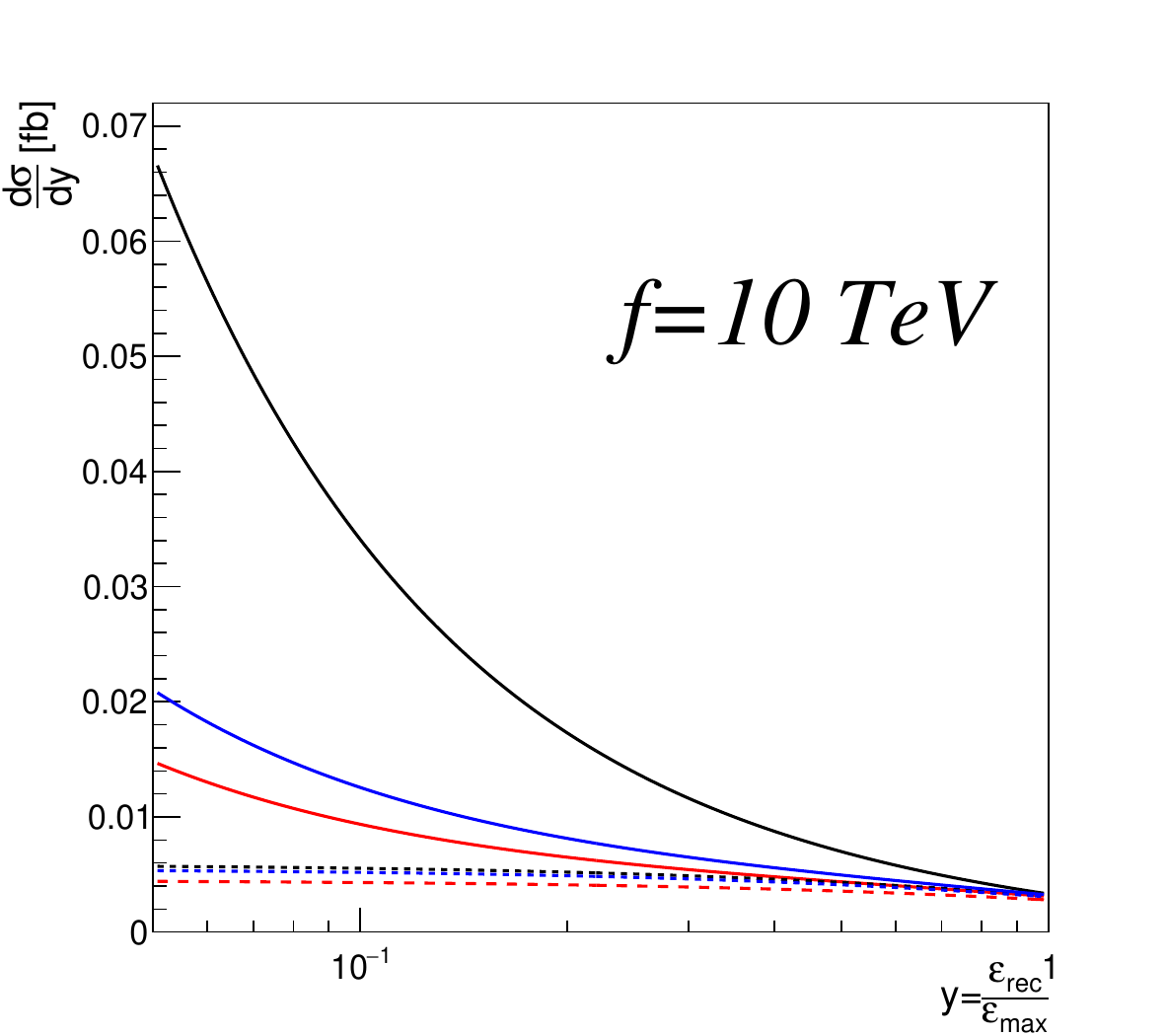}
\caption{}
\label{fig1:sigmaAr_sub2}
\end{subfigure}
\caption{Differential cross section of the LDCP-Ar scattering as a function of $y$ for $f=5\,\mbox{TeV}$, $c_V^{\chi}=0.5$
and $\varepsilon_H=-0.1$ {\it (a)} as well as for $f=10\,\mbox{TeV}$, $c_V^{\chi}=0.5$ and
$\varepsilon_H=-0.1$ {\it (b)}. Red, blue and black lines are associated with $m_{\chi}=200\,\mbox{GeV}$,
$m_{\chi}=500\,\mbox{GeV}$ and $m_{\chi}=1000\,\mbox{GeV}$, respectively. Dashed and solid lines correspond to $\mu_{\chi}=0$ and
$\mu_{\chi}=3\cdot \mu_{\chi}^{\text{exp}}(m_{\chi})$.}
\label{fig3.0}
\end{figure}

The results of our study of the differential cross sections of the LDCP--Xe and LDCP--Ar scattering are shown in
Fig.~\ref{fig2.0} and \ref{fig3.0} respectively. Since in the xenon (argon) based experiments it is rather problematic
to detect a signal if $\varepsilon_{rec}\lesssim 2-3\,\mbox{KeV}$ in our analysis we set lower bounds $y_{min}$ on the intervals
of variations of $y$. These limits are $y_{min}\simeq 10^{-2}$ and $y_{min}\simeq 4\cdot 10^{-2}$ in the cases of LDCP--Xe and
LDCP--Ar scattering respectively. To illustrate our findings we use some of the benchmark scenarios listed in Table~\ref{tab:BM}.
These scenarios are associated with $c^{\chi}_V=0.5$, $\varepsilon_{H}=-0.1$ as well as $f=5\,\mbox{TeV}$ and $f=10\,\mbox{TeV}$.
In order to ensure the phenomenological viability of these scenarios we set $\xi=0.1$.
Since the experimental bound on the magnetic dipole moment $\mu^{exp}_{\chi}$ of the LDCP implies that the DM fermion $\chi$ constitutes
the entirety of the DM, this limit becomes somewhat weaker in the case $\xi \le 1$., i.e
\begin{equation}
\mu_{\chi} < \dfrac{\mu^{exp}_{\chi}}{\sqrt{\xi}}\,.
\label{34}
\end{equation}

In Figs.~\ref{fig2.0} and \ref{fig3.0} we compare scenarios with $\mu_{\chi}=0$ and $\mu_{\chi}\simeq 3\cdot \mu^{exp}_{\chi}(m_{\chi})$
for two different values of the compositeness scale ($f\simeq 5\,\mbox{TeV}$ and $f\simeq 10\,\mbox{TeV}$) and
the LDCP masses equal to $200\,\mbox{GeV}$, $500\,\mbox{GeV}$ and $1\,\mbox{TeV}$.
Figs.~\ref{fig2.0}a and \ref{fig3.0}a demonstrate that for relatively low scales $f \lesssim 10\,\mbox{TeV}$
all terms, which are proportional to $\mu_{\chi}$, give only a minor contribution to the differential cross section (\ref{32}).
Nevertheless as $f$ increases, the Higgs and $Z$ boson contributions to the LDCP–-nucleus scattering amplitude decrease,
so that the electromagnetic contribution to the differential cross section (\ref{eq3.12}) is getting more substantial for larger $f$.
From Figs.~\ref{fig2.0}b and \ref{fig3.0}b one can see that in the composite Higgs models with approximate $U(1)$ symmetry
such contribution becomes sizable for $y\ll 1$ if $f \gtrsim 10\,\mbox{TeV}$ and $c_V^{\chi}$ is considerably smaller than unity.
Furthermore, the growth of the differential cross section at low $y$ is more apparent in the part of the parameter space where
the contributions of the diagrams with t-channel $Z$--boson exchange and t-channel exchange of the
SM--like Higgs partially cancel each other. This corresponds to the scenarios with $c_V^{\chi}\simeq 0.5$ and $\varepsilon_H=-0.1$.
The enhancement of the differential cross section (\ref{eq3.12}) for $y\ll 1$ is greater for the argon target.
This is because in the case of argon the first constant term in Eq.~(\ref{eq3.12}) is suppressed by a factor $\left(A_{Ar}/A_{Xe}\right)^2$
as compared with the similar one in the differential cross section of the LDCP--Xe scattering.

\subsection{Event rate}

Eq.~(\ref{eq3.12}) clearly indicates that for substantial value of the magnetic moment of the LDCP and sufficiently small $y$
the differential cross section (\ref{32}) should grow with decreasing $y$ or nuclear recoil energy $\varepsilon_{rec}$.
In principle such growth may permit to distinguish the DM fermions with non--zero $\mu_{\chi}$ from other types of
dark matter particles that don't have similar electromagnetic properties. In this context it is truly remarkable that
xenon based experiment LZ strengthened bounds on the SI DM--nucleon scattering cross section by factor of $10$ within last few years.
If this trend continues there is a chance that the DM signal is going to be observed soon. The aim of our study here
is to explore whether electromagnetic properties of the DM fermions in the CHMs can be probed in such direct detection
experiments.

To assess the observability of the growth rate of the differential cross section (\ref{32}) at low recoil energies $\varepsilon_{rec}$
(or $y$) we investigate the event rate of LDCP--Xe (Ar) scattering ($\chi+A\to \chi+A$) in direct detection experiments.
The differential event rate can be evaluated via \cite{Hambye:2021xvd}:
\begin{equation}
\frac{dR_{A}(\mu_{\chi})}{d\varepsilon_{rec}}=\epsilon (\varepsilon_{rec}) N_T n_{\chi}
\int d^3 {\bf v}_{\chi} f({\bf v}_{\chi})\dfrac{m_T}{2 M_r^2 |{\bf v}_{\chi}|} \left(\frac{d\sigma}{dy}\right)\,
\Theta(|{\bf v}_{\chi}| - v_{min})\,,
\label{35}
\end{equation}
where detector efficiency $\epsilon (\varepsilon_{rec})$ is approximated by $\Theta(\varepsilon_{rec}-\varepsilon_{\text{th}})$, $\varepsilon_{\text{th}}\simeq 2\,\mbox{KeV}$, $N_T$ is the total number of target nuclei, $n_{\chi}$ is the local LDCP number density,
i.e. $n_{\chi}=\xi \rho_{0}/m_{\chi}$, and $\rho_{0}=0.4\,\mbox{GeV}/\mbox{cm}^3$\,. Here $v_{min}$ is the minimal velocity that can give
rise to a given $\varepsilon_{rec}$, i.e.
\begin{equation}
v_{min}=\sqrt{\dfrac{m_T \varepsilon_{rec}}{2 M_r^2}}\,.
\label{36}
\end{equation}
In Eq.~(\ref{35}) the LDCP velocity distribution $f({\bf v}_{\chi})$ is parametrized by a truncated Maxwellian distribution
in the frame of the Galaxy and then boosted to the Earth frame so that
\begin{equation}
f({\bf v}_{\chi})=\frac{1}{N_0(v_0, v_{esc})}\exp\left[-\dfrac{\tilde{v}^2}{v_0^2}\right]\,\Theta(v_{esc}-\tilde{v})\,,
\label{37}
\end{equation}
where $\tilde{v}=|{\bf v}_{\chi}+{\bf v}_{E}|$, $|{\bf v}_{E}|\approx 240\,\text{km}/\text{s}$ is the Earth velocity with respect to the Galaxy,
$v_{esc}\approx 550\,\text{km}/\text{s}$ is the escape velocity of the Galaxy and $v_0\simeq 220\,\text{km}/\text{s}$ is the mean velocity
of the Maxwellian distribution. In Eq.~(\ref{37}) the factor $N_0(v_0, v_{esc})$ is introduced to normalize
$f({\bf v}_{\chi})$ so that $\int d^3 {\bf v}_{\chi} f({\bf v}_{\chi})=1$ \cite{Hambye:2021xvd}.

The total event rate in direct detection experiments can be estimated as
\begin{equation}
R^A_{tot}(\mu_{\chi})\simeq\int_{\varepsilon_{\text{th}}}^{\varepsilon_0} d\varepsilon_{rec}
\frac{dR_{A}(\mu_{\chi})}{d\varepsilon_{rec}}\,.
\label{38}
\end{equation}
In principle, the upper limit $\varepsilon_0$ of an integral (\ref{38}) should be set equal to infinity.
However from Eq.~(\ref{f_SI}) as well as from Figs.~\ref{fig2.0} and \ref{fig3.0} one can see
that $F_{SI}({\bf q}^2)$ and the differential cross section (\ref{32}) decrease substantially
with the growth of the recoil energy $\varepsilon_{rec}$. It is easy to check that the total
event rate $R^A_{tot}$ remains approximately the same if instead of infinity $\varepsilon_0$
is set to be equal to $80-100\,\mbox{KeV}$. In our analysis we fix $\varepsilon_{0}\simeq 100\,\mbox{KeV}$.

\begin{table}[htbp]
\centering
\caption{SD fraction of the total LDCP--Xe event rate (\(N\)) for $c^{\chi}_V=0.5$,
\(c_{PV}^\chi=0.5\) and \(\mu_\chi=3\,\mu_\chi^{\rm exp}(m_\chi)\), computed with the simplified SD form factor (\(N_{SD}/N\))
and with the nuclear response functions of Ref.~\cite{Fitzpatrick:2012ix} (\(N^{\rm NR}_{SD}/N\)). The last two columns show
the corresponding values of observables \(\Delta_{Xe}\) and \(\Delta^{NR}_{Xe}\). For each benchmark scenario the computed
values of \(\Delta_{Ar}\) and $r$ are also included.}
\label{tab:fractions}
\begin{tabular}{ccccccccc}
\toprule
\(m_\chi\) [GeV] & \(f\) [TeV] & \(\varepsilon_H\) & \(N_{SD}/N\) [\%] & \(N^{\rm NR}_{SD}/N\) [\%] & $r$ & \(\Delta_{Ar}\) &
\(\Delta_{Xe}\) & \(\Delta^{NR}_{Xe}\)\\
\midrule
	& 10 & 0    & 0.3 & 0.25& 0.24 & 0.108 & 0.190  & 0.190  \\
200	& 10 & -0.1 & 1.2 & 0.9 & 0.072 & 0.175 & 0.218  & 0.218 \\
	& 15 & 0    & 1.3 & 0.9 & 0.059 & 0.179 & 0.227  & 0.227  \\
	& 15 & -0.1 & 5.5 & 4.0 & 0.0132 & 0.356 & 0.386  & 0.392  \\
\midrule
    &  5 & -0.1 & 0.1 & 0.08 & 1.01 & 0.079 & 0.173 & 0.172 \\
	& 10 & 0    & 0.5 & 0.4  & 0.12 & 0.113 & 0.188  & 0.186  \\
500	& 10 & -0.1 & 1.7 & 1.3  & 0.036& 0.196 & 0.228  & 0.229 \\
	& 15 & 0    & 1.9 & 1.4  & 0.030& 0.201 & 0.240  & 0.242  \\
	& 15 & -0.1 & 6.4 & 4.7  & 0.0088& 0.357 & 0.407  & 0.414  \\
\midrule
    &  5 & -0.1 & 0.17& 0.14 & 0.41  & 0.083 & 0.173 & 0.173 \\
	& 10 & 0    & 1.0 & 0.8 & 0.051 & 0.146 & 0.205  & 0.205  \\
1000& 10 & -0.1 & 2.9 & 2.2 & 0.018 & 0.252 & 0.273  & 0.275  \\
	& 15 & 0    & 3.2 & 2.4 & 0.015 & 0.257 & 0.289  & 0.292  \\
	& 15 & -0.1 & 7.1 & 5.3 & 0.0071 & 0.363 & 0.433  & 0.441  \\
\bottomrule
\end{tabular}
\end{table}

Again, we use benchmark scenarios specified in Table~\ref{tab:BM} to analyse the event rate.
Since Figs.~\ref{fig2.0}a and \ref{fig3.0}a indicate that for $f \lesssim 10\,\mbox{TeV}$ the electromagnetic
interaction gives only a minor contribution to the differential cross section (\ref{32}), so that it is going to be
rather challenging to probe the electromagnetic properties of the dark matter particles, here we therefore examine
scenarios with larger compositeness scales, specifically $f \simeq 10\,\text{TeV}$ and $f \simeq 15\,\text{TeV}$.
The results shown in Table \ref{tab:dsigma} indicate that the SD part of the differential cross section (\ref{32})
is rather small. Therefore we calculate the contribution of this part to the event rate separately.
In Table \ref{tab:fractions} the relative contributions of the SD part of the differential cross section of the LDCP--nucleus scattering
to the total event rate computed within the simplified approach and in the framework of response-function formalism
are presented. One can see that the contribution of SI part to the event rate is more than an order of magnitude
larger than the contribution of SD one. The smallness of the contribution of the SD part simplifies the
comparison of the computed event rate with the current experimental limits $R_0$. For each benchmark scenario
$R_0$ can be identified with the event rate $R^{Xe}_{tot}(0)$ corresponding to the maximal possible value
of the spin--independent DM--nucleon scattering cross section $\sigma^N_{SI\,\text{exp}}(m_{\chi})$
and $\sigma^N_{SD}=0$. In Table \ref{tab:fractions} for each benchmark scenario the ratio
\begin{equation}
r=R^A_{tot}(3\cdot \mu^{exp}_{\chi}(m_{\chi})/R_0
\label{381}
\end{equation}
is given assuming that $\xi=0.1$\,. One can see that almost all values of $r$ are
considerably smaller than unity so that our benchmark scenarios have not been ruled out.
From Table \ref{tab:fractions} it becomes clear that the accuracy of our calculation of the event rate cannot exceed a few percents.

\begin{figure}[htbp]
\centering
\begin{subfigure}[b]{0.47\textwidth}
\centering
\includegraphics[width=\textwidth]{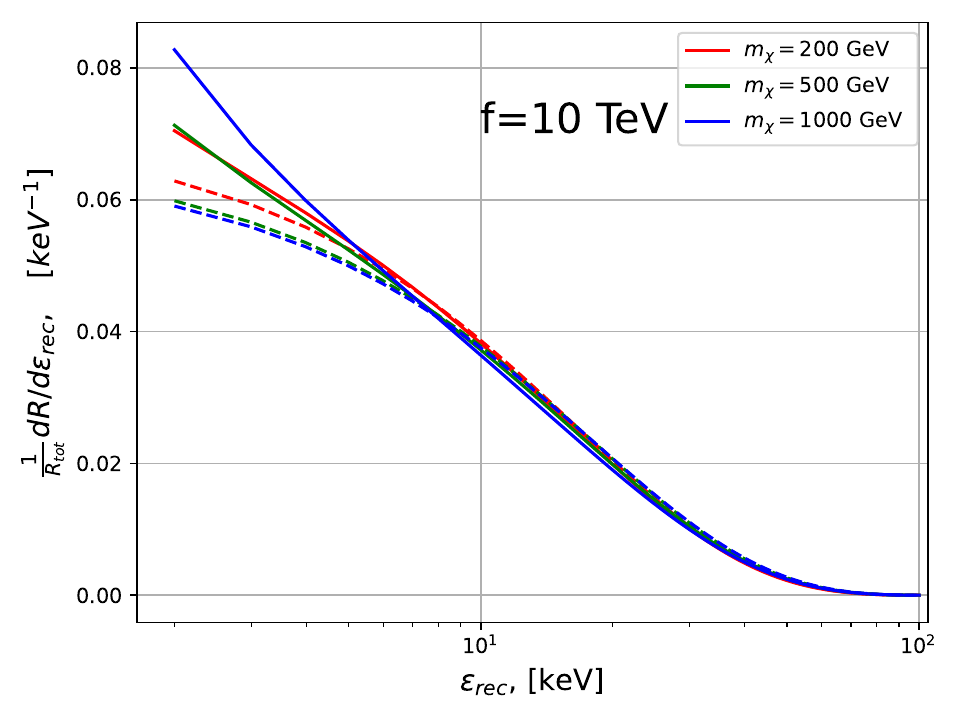}
\caption{}
\label{fig1:sub1}
\end{subfigure}
\hfill  
\begin{subfigure}[b]{0.47\textwidth}
\centering
\includegraphics[width=\textwidth]{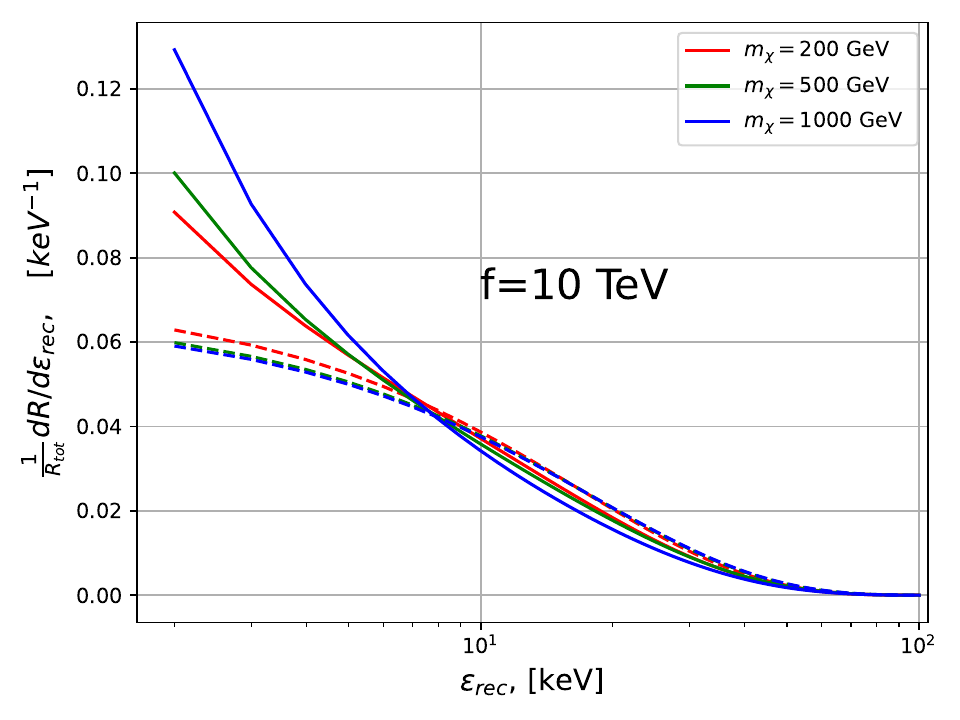}
\caption{}
\label{fig1:sub2}
\end{subfigure}
\vspace{0.2cm}
\begin{subfigure}[b]{0.47\textwidth}
\centering
\includegraphics[width=\textwidth]{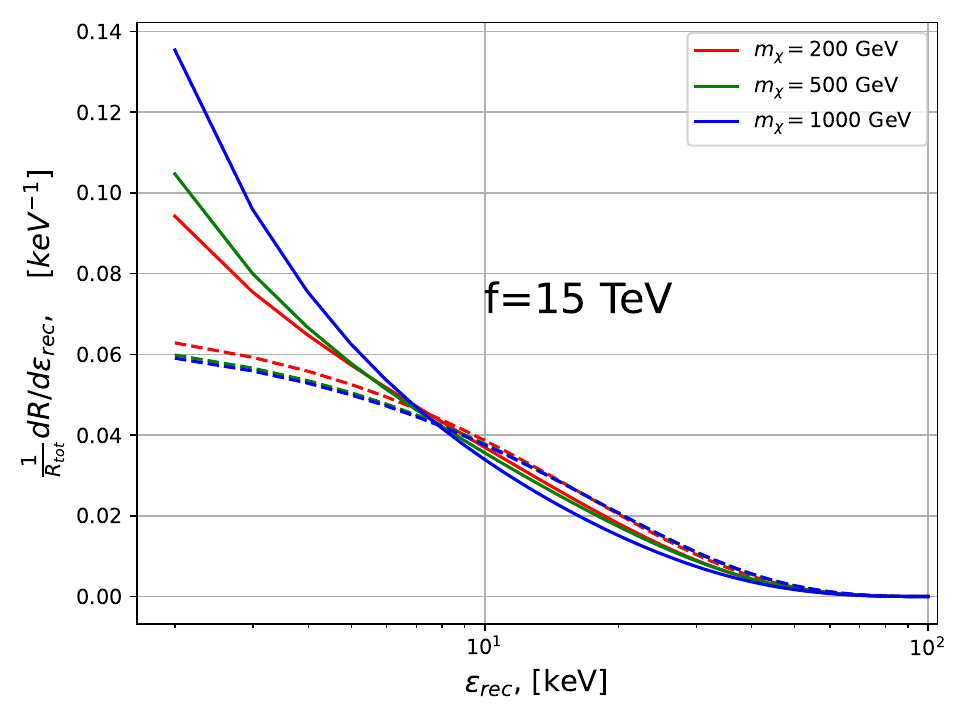}
\caption{}
\label{fig1:sub3}
\end{subfigure}
\hfill
\begin{subfigure}[b]{0.47\textwidth}
\centering
\includegraphics[width=\textwidth]{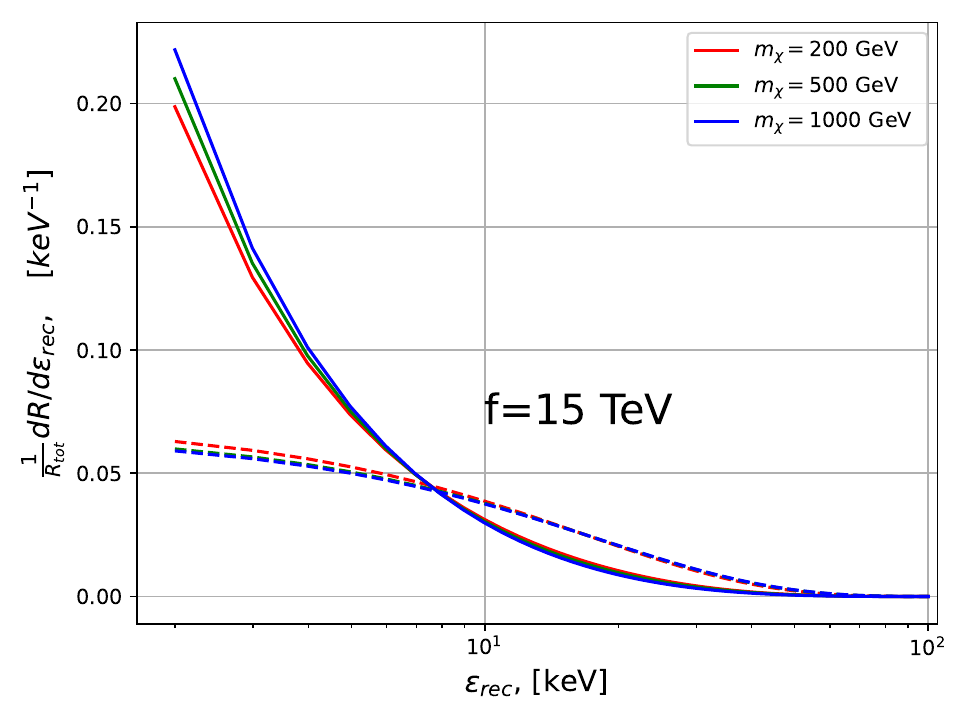}
\caption{}
\label{fig1:sub4}
\end{subfigure}
\caption{The normalized differential event rate
$\dfrac{1}{R^{Xe}_{tot}}\dfrac{dR_{Xe}}{d\varepsilon_{rec}}$ of the LDCP-Xe scattering
as a function of recoil energy of Xe nuclei for
$f= 10\,\text{TeV}$, $c_V^{\chi}=0.5$ and $\varepsilon_{H}=0$ {\it (a)},
$f= 10\,\text{TeV}$, $c_V^{\chi}=0.5$ and $\varepsilon_{H}=-0.1$ {\it (b)},
$f= 15\,\text{TeV}$, $c_V^{\chi}=0.5$ and $\varepsilon_{H}=0$ {\it (c)} as well as
$f= 15\,\text{TeV}$, $c_V^{\chi}=0.5$ and $\varepsilon_{H}=-0.1$ {\it (d)}.
Red, green and blue lines are associated with $m_{\chi}=200\,\mbox{GeV}$,
$m_{\chi}=500\,\mbox{GeV}$ and $m_{\chi}=1000\,\mbox{GeV}$, respectively.
Solid and dashed lines correspond to $\mu_{\chi}=3\cdot \mu_{\chi}^{\text{exp}}(m_{\chi})$ and
$\mu_{\chi}=0$.}
\label{fig4.0}
\end{figure}

\begin{figure}[htbp]
\centering
\begin{subfigure}[b]{0.47\textwidth}
\centering
\includegraphics[width=\textwidth]{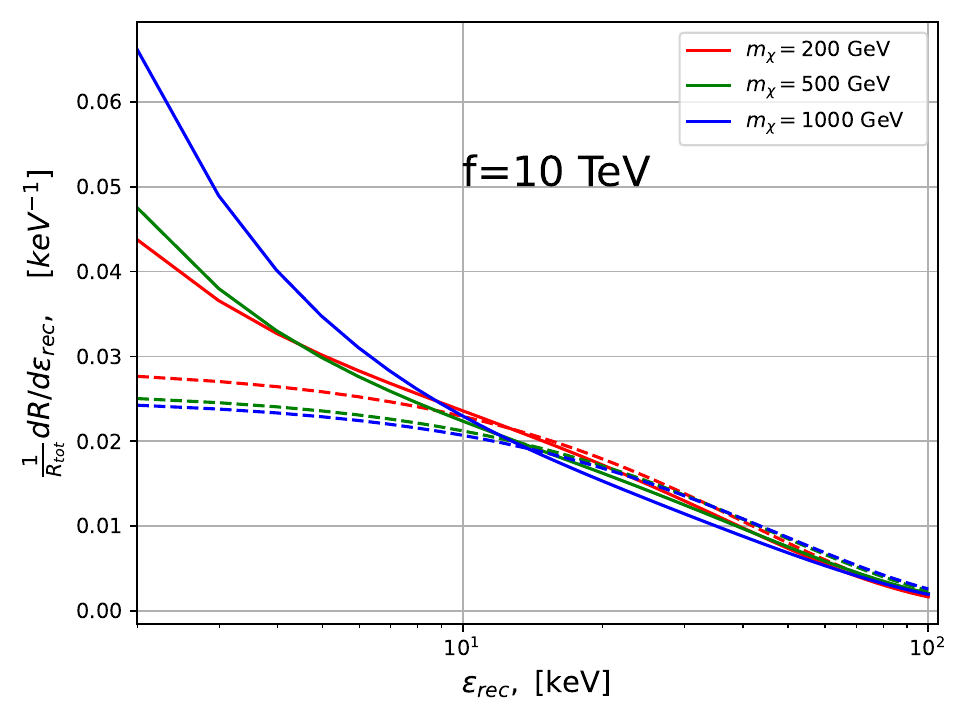}
\caption{}
\label{fig1:sub1_ar}
\end{subfigure}
\hfill  
\begin{subfigure}[b]{0.47\textwidth}
\centering
\includegraphics[width=\textwidth]{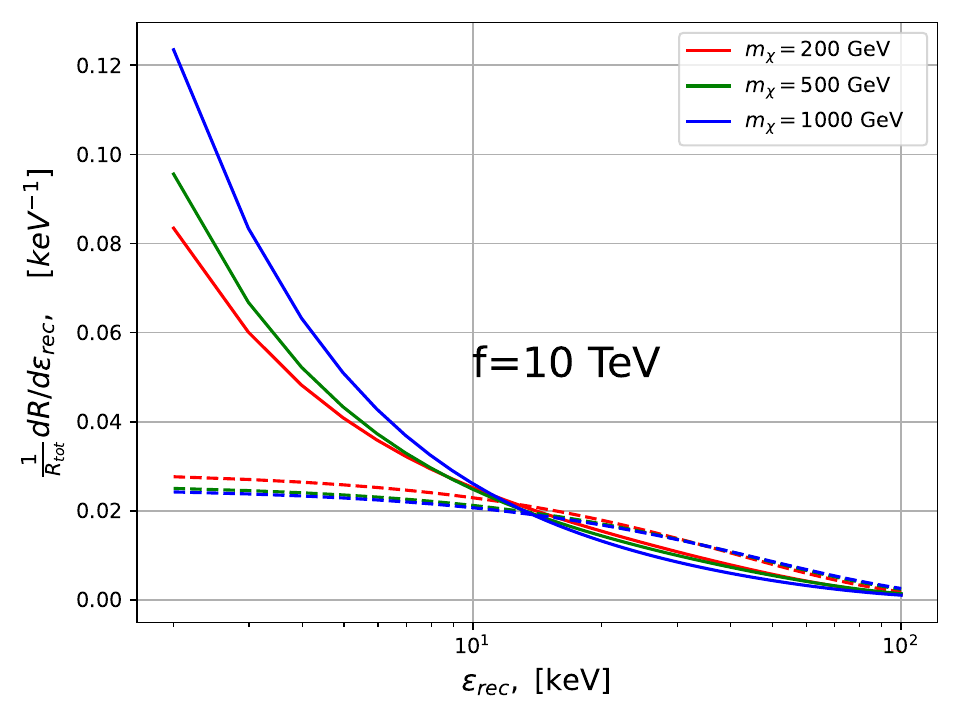}
\caption{}
\label{fig1:sub2_ar}
\end{subfigure}
\vspace{0.2cm}
\begin{subfigure}[b]{0.47\textwidth}
\centering
\includegraphics[width=\textwidth]{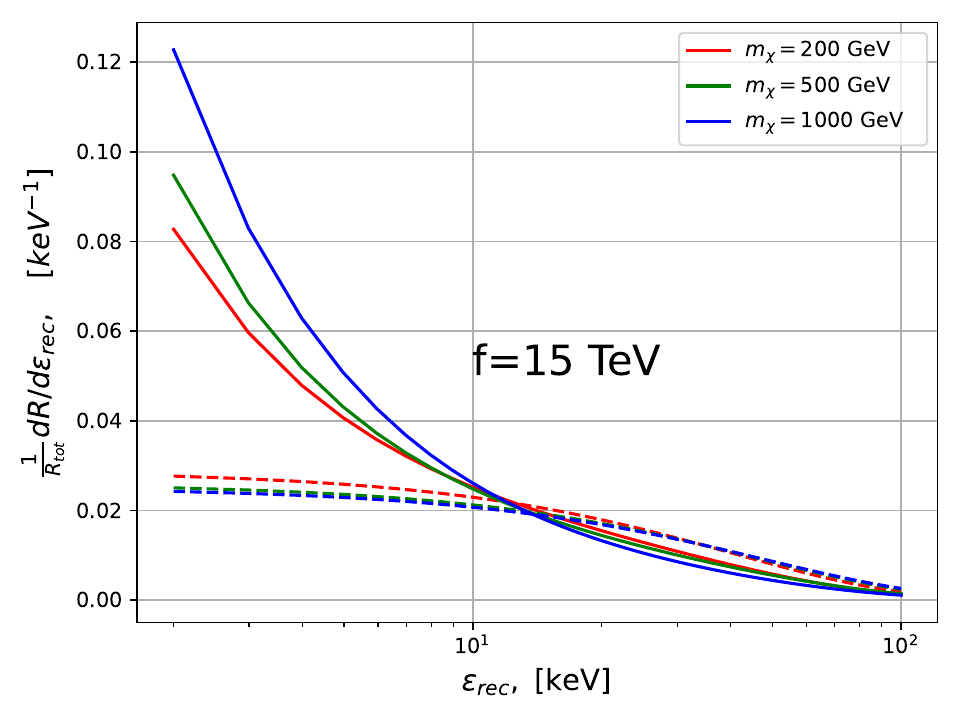}
\caption{}
\label{fig1:sub3_ar}
\end{subfigure}
\hfill
\begin{subfigure}[b]{0.47\textwidth}
\centering
\includegraphics[width=\textwidth]{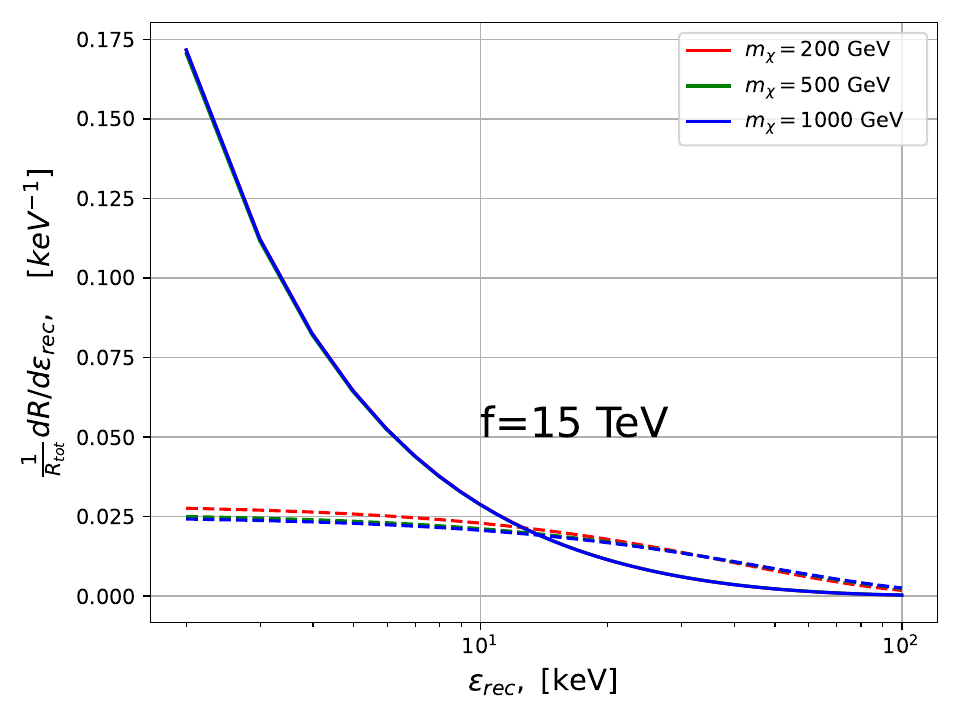}
\caption{}
\label{fig1:sub4_ar}
\end{subfigure}
\caption{
The normalized differential event rate
$\dfrac{1}{R^{Ar}_{tot}}\dfrac{dR_{Ar}}{d\varepsilon_{rec}}$ of the LDCP-Ar scattering
as a function of recoil energy of Ar nuclei for
$f= 10\,\text{TeV}$, $c_V^{\chi}=0.5$ and $\varepsilon_{H}=0$ {\it (a)},
$f= 10\,\text{TeV}$, $c_V^{\chi}=0.5$ and $\varepsilon_{H}=-0.1$ {\it (b)},
$f= 15\,\text{TeV}$, $c_V^{\chi}=0.5$ and $\varepsilon_{H}=0$ {\it (c)} as well as
$f= 15\,\text{TeV}$, $c_V^{\chi}=0.5$ and $\varepsilon_{H}=-0.1$ {\it (d)}.
Red, green and blue lines are associated with $m_{\chi}=200\,\mbox{GeV}$,
$m_{\chi}=500\,\mbox{GeV}$ and $m_{\chi}=1000\,\mbox{GeV}$, respectively.
Solid and dashed lines correspond to $\mu_{\chi}=3\cdot \mu_{\chi}^{\text{exp}}(m_{\chi})$ and
$\mu_{\chi}=0$.}
\label{fig5.0}
\end{figure}

In Figs.~\ref{fig4.0} and \ref{fig5.0} the normalized differential event rate, i.e.
\begin{equation}
\mathcal{G}_A(\varepsilon_{rec})=\dfrac{1}{R^{A}_{tot}}\dfrac{dR_{A}}{d\varepsilon_{rec}}\,,
\label{381}
\end{equation}
is explored as a function of the recoil energy of nuclei for the LDCP--Xe and LDCP--Ar scattering.
We compare scenarios with $\mu_{\chi}=0$ and $\mu_{\chi}\simeq 3\cdot \mu^{exp}_{\chi}(m_{\chi})$
for three different values of the LDCP masses $m_{\chi}=200\,\mbox{GeV}$, $500\,\mbox{GeV}$ and $1\,\mbox{TeV}$.
In general for $f \simeq 10\,\mbox{TeV}$ the $Z$ boson contribution
to the differential cross section of the LDCP--Xe scattering is still too large as compared with the contribution of electromagnetic
interaction (see Fig.~\ref{fig4.0}a). Nonetheless there is a part of the parameter space near $c_V^{\chi}\simeq 0.5$ and
$\varepsilon_H=-0.1$ where the SM--like Higgs and $Z$ boson contributions to the LDCP--Xe scattering amplitude partially cancel
each other\footnote{Such partial cancellation of the SM-like Higgs and $Z$ boson
contributions to the SI part of the DM–-nucleus scattering amplitude takes place for all benchmark scenarios with $\varepsilon_H=-0.1$.
The degree of this cancellation is given by $\sqrt{\dfrac{\sigma^N_{SI}(\varepsilon_H=-0.1)}{\sigma^N_{SI}(\varepsilon_H=0)}}$.}.
This results in the noticeable enhancement of the normalized differential event rate at low recoil energies
$\varepsilon_{rec}\lesssim 10\,\mbox{KeV}$ (see Fig.~\ref{fig4.0}b) that might be possible to detect in the future.
When $f \simeq 15\,\mbox{TeV}$ the combined Higgs and $Z$ boson contribution to the differential cross section of the LDCP--Xe scattering
reduces by factor $5$ or even stronger. Therefore the growth of the normalized differential event rate at low recoil energies becomes
considerably larger (see Figs.~\ref{fig4.0}c and \ref{fig4.0}d). The results presented in Fig.~\ref{fig5.0}
demonstrate that the corresponding enhancement of the normalized differential event rate tends to bigger in the case of argon
nuclei because $A_{Ar}\ll A_{Xe}$.

\begin{figure}[htbp]
\centering
\begin{subfigure}[b]{0.47\textwidth}
\centering
\includegraphics[width=\textwidth]{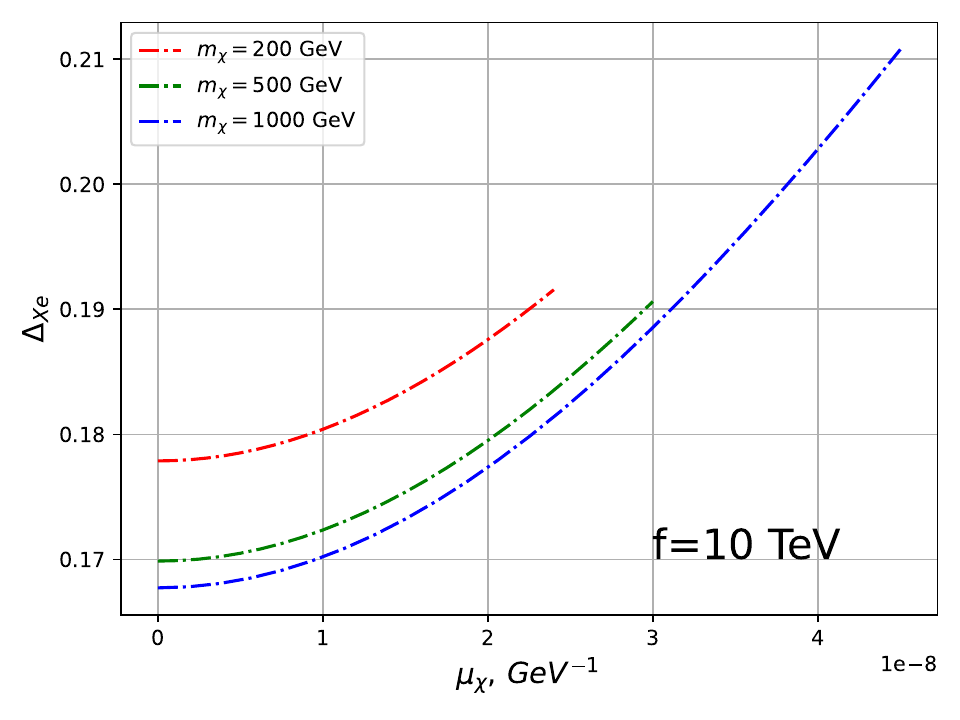}
\caption{}
\label{fig1:sub1d}
\end{subfigure}
\hfill  
\begin{subfigure}[b]{0.47\textwidth}
\centering
\includegraphics[width=\textwidth]{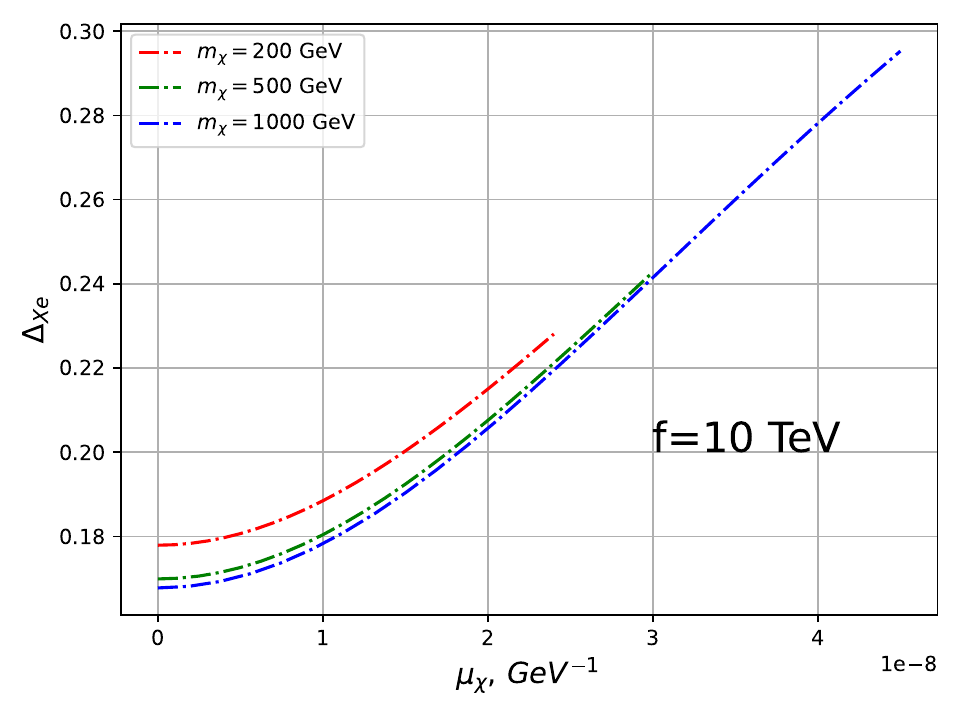}
\caption{}
\label{fig1:sub2d}
\end{subfigure}
\vspace{0.2cm}
\begin{subfigure}[b]{0.47\textwidth}
\centering
\includegraphics[width=\textwidth]{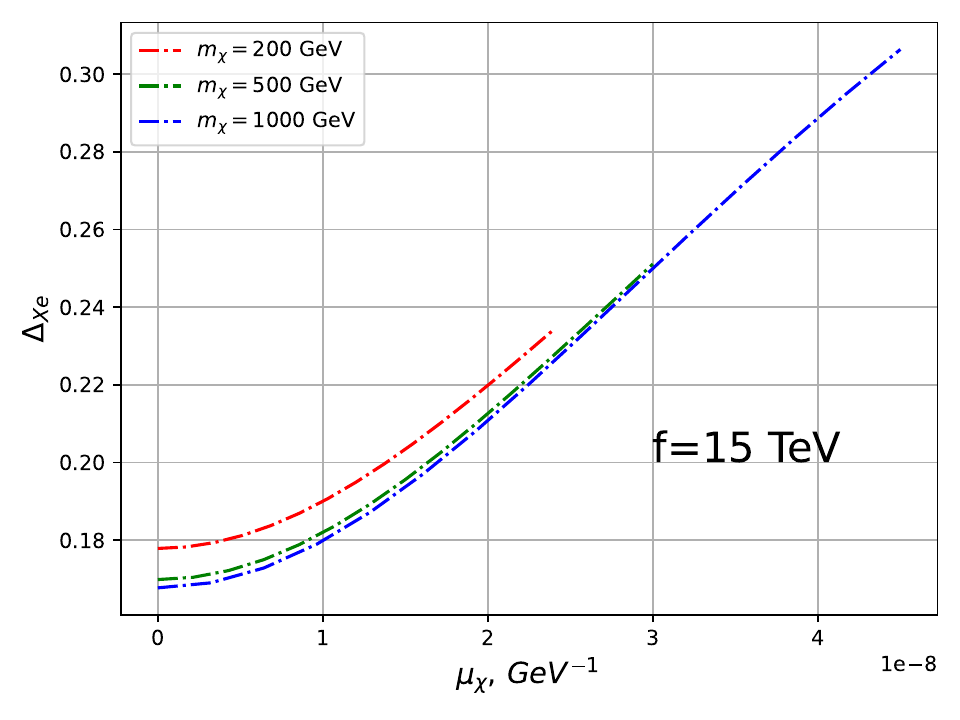}
\caption{}
\label{fig1:sub3d}
\end{subfigure}
\hfill
\begin{subfigure}[b]{0.47\textwidth}
\centering
\includegraphics[width=\textwidth]{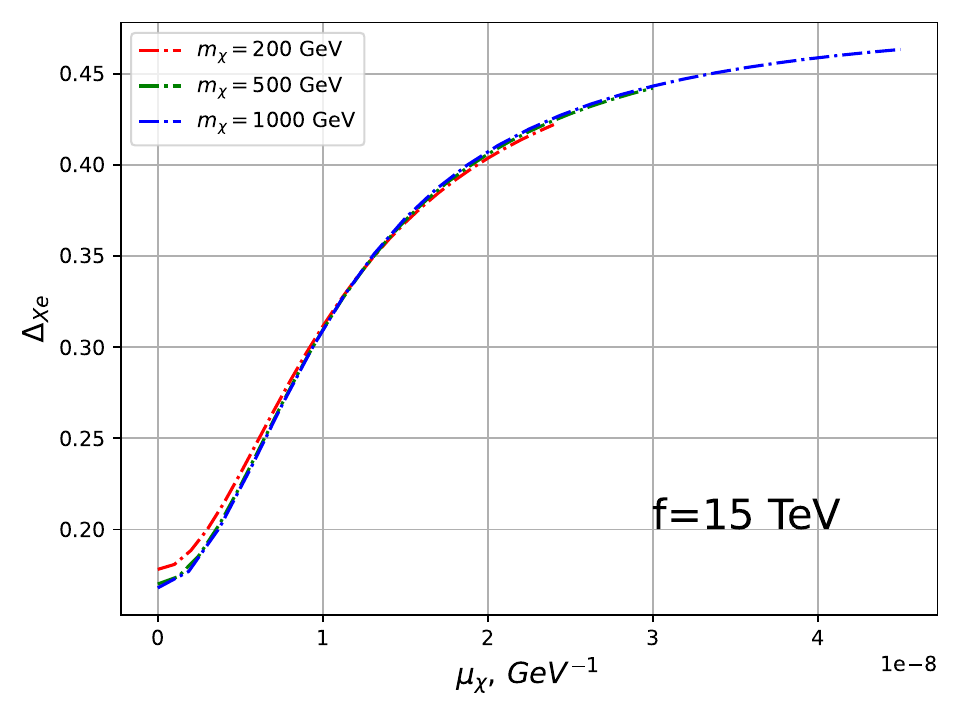}
\caption{}
\label{fig1:sub4d}
\end{subfigure}
\caption{Dependence of $\Delta_{Xe}(\mu_{\chi})$ on the LDCP magnetic moment $\mu_{\chi}$ for
$f= 10\,\text{TeV}$, $c_V^{\chi}=0.5$ and $\varepsilon_{H}=0$ {\it (a)},
$f= 10\,\text{TeV}$, $c_V^{\chi}=0.5$ and $\varepsilon_{H}=-0.1$ {\it (b)},
$f= 15\,\text{TeV}$, $c_V^{\chi}=0.5$ and $\varepsilon_{H}=0$ {\it (c)} as well as
$f= 15\,\text{TeV}$, $c_V^{\chi}=0.5$ and $\varepsilon_{H}=-0.1$ {\it (d)}.
Red, green and blue dashed--dotted lines are associated with $m_{\chi}=200\,\mbox{GeV}$,
$m_{\chi}=500\,\mbox{GeV}$ and $m_{\chi}=1000\,\mbox{GeV}$, respectively.
}
\label{fig6.0}
\end{figure}

\begin{figure}[htbp]
\centering
\begin{subfigure}[b]{0.47\textwidth}
\centering
\includegraphics[width=\textwidth]{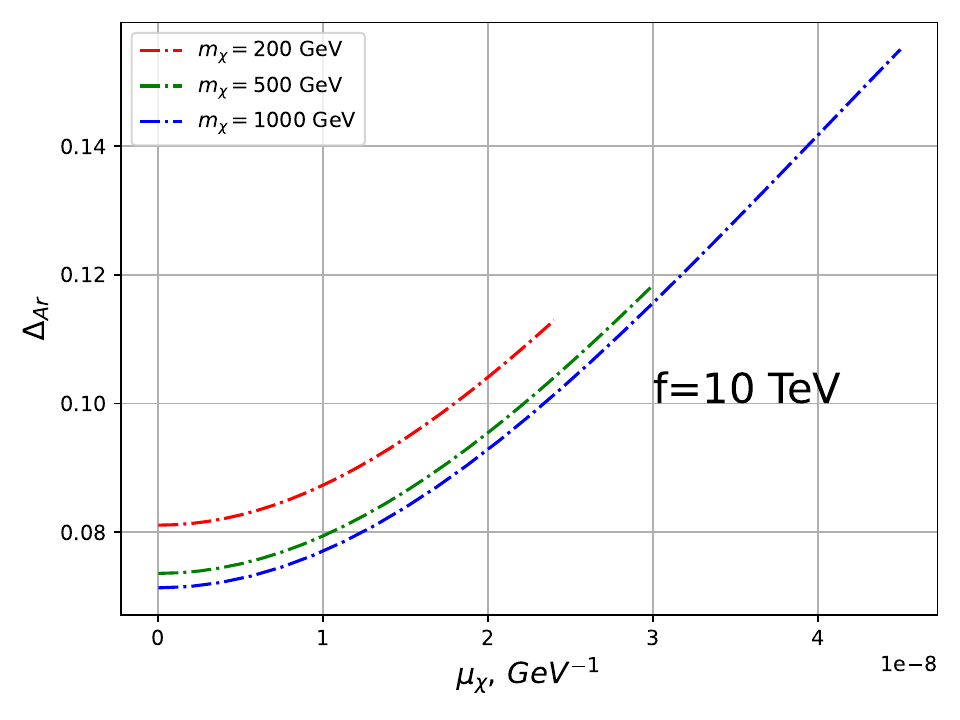}
\caption{}
\label{fig1:sub1d_Ar}
\end{subfigure}
\hfill  
\begin{subfigure}[b]{0.47\textwidth}
\centering
\includegraphics[width=\textwidth]{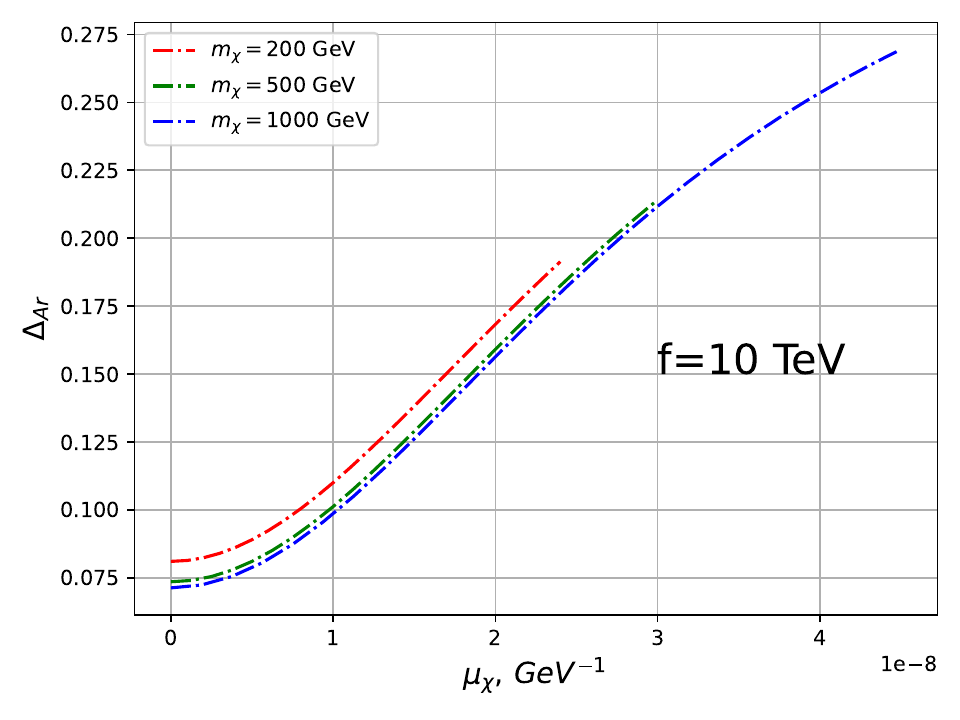}
\caption{}
\label{fig1:sub2d_Ar}
\end{subfigure}
\vspace{0.2cm}
\begin{subfigure}[b]{0.47\textwidth}
\centering
\includegraphics[width=\textwidth]{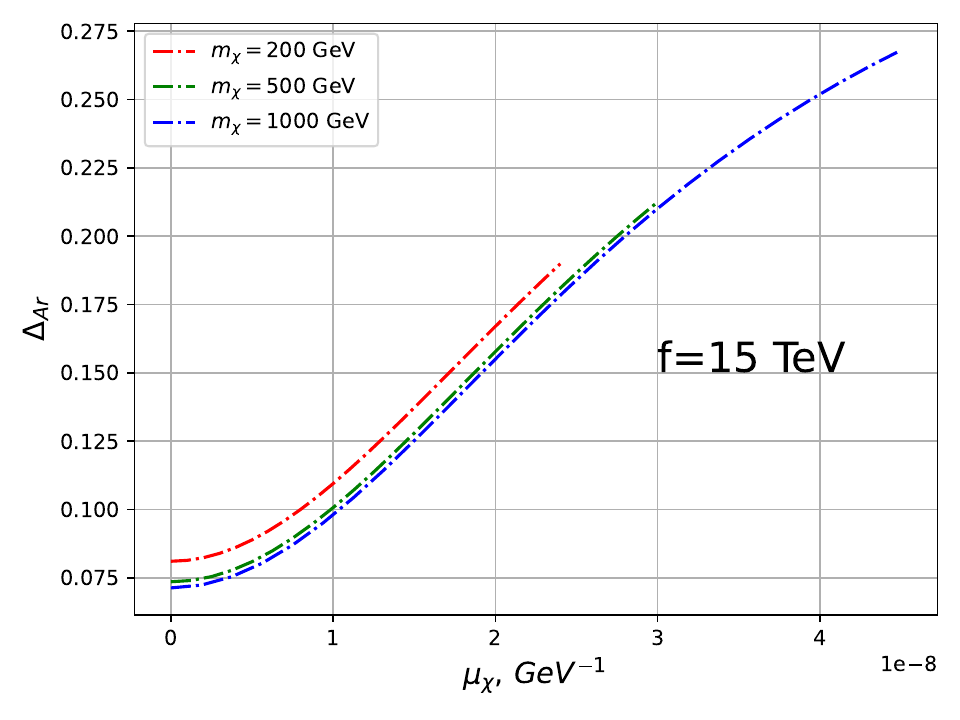}
\caption{}
\label{fig1:sub3d_Ar}
\end{subfigure}
\hfill
\begin{subfigure}[b]{0.47\textwidth}
\centering
\includegraphics[width=\textwidth]{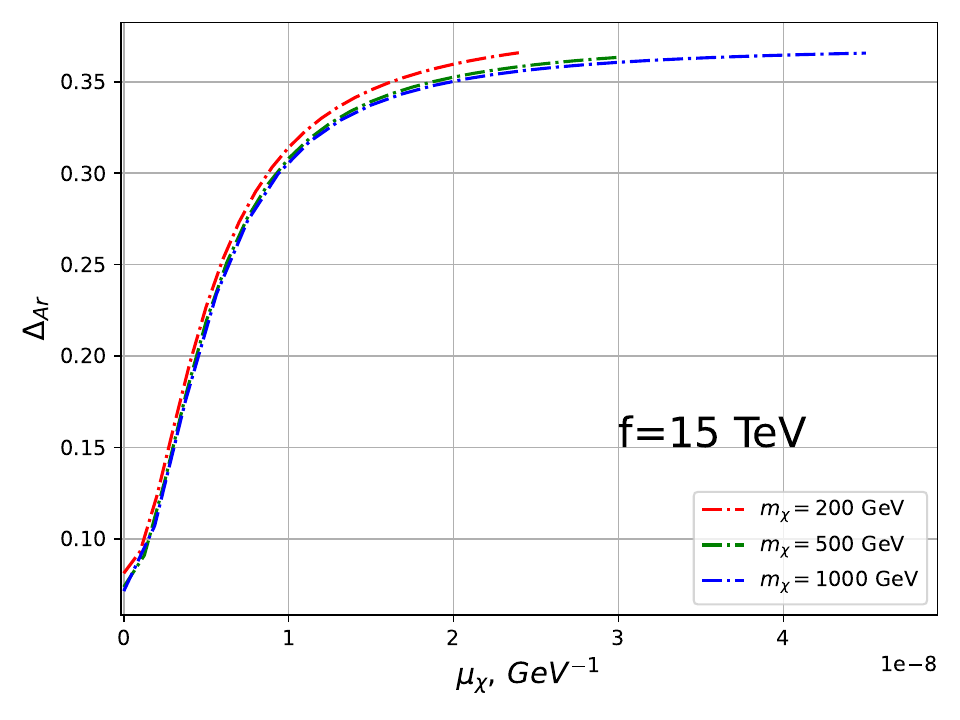}
\caption{}
\label{fig1:sub4d_Ar}
\end{subfigure}
\caption{Dependence of $\Delta_{Ar}(\mu_{\chi})$ on the LDCP magnetic moment $\mu_{\chi}$ for
$f= 10\,\text{TeV}$, $c_V^{\chi}=0.5$ and $\varepsilon_{H}=0$ {\it (a)},
$f= 10\,\text{TeV}$, $c_V^{\chi}=0.5$ and $\varepsilon_{H}=-0.1$ {\it (b)},
$f= 15\,\text{TeV}$, $c_V^{\chi}=0.5$ and $\varepsilon_{H}=0$ {\it (c)} as well as
$f= 15\,\text{TeV}$, $c_V^{\chi}=0.5$ and $\varepsilon_{H}=-0.1$ {\it (d)}.
Red, green and blue dashed--dotted lines correspond to $m_{\chi}=200\,\mbox{GeV}$,
$m_{\chi}=500\,\mbox{GeV}$ and $m_{\chi}=1000\,\mbox{GeV}$, respectively.}
\label{fig7.0}
\end{figure}

The enhancement of the event rate at low recoil energies caused by the electromagnetic interaction of the LDCP with nuclei might be
observed only if there is a substantial fraction of such events at small $\varepsilon_{rec}$.
From Fig.~\ref{fig4.0} and \ref{fig5.0} one can see that the sizable deviation of the differential event rates
calculated for $\mu_{\chi}=0$ and $\mu_{\chi}\simeq 3\cdot \mu^{exp}_{\chi}(m_{\chi})$ takes place when
$\varepsilon_{rec}\lesssim 10\,\mbox{KeV}$. To quantify such deviation we define the following quantity:
\begin{equation}
\Delta_{A}(\mu_{\chi})=\dfrac{1}{R^A_{tot}(\mu_{\chi})}\int_{\varepsilon_{\text{th}}}^{5\,\text{KeV}} d\varepsilon_{rec}
\frac{dR_{A}(\mu_{\chi})}{d\varepsilon_{rec}}\,,\qquad\qquad \Delta_{A}=\Delta_{A}(3\cdot \mu^{exp}_{\chi}(m_{\chi}))\,.
\label{39}
\end{equation}
For each benchmark scenario we compute $\Delta_{Xe}(\mu_{\chi})$ and $\Delta_{Xe}$
associated with the LDCP--Xe scattering as well as $\Delta_{Ar}(\mu_{\chi})$ and $\Delta_{Ar}$
corresponding to the LDCP--Ar scattering. The results of our numerical calculations are shown in Table~\ref{tab:fractions}
as well as Figs.~\ref{fig6.0} and \ref{fig7.0}.
For the LDCP--Xe scattering the values of observable (\ref{39}) in Table~\ref{tab:fractions} is computed within the simplified approach
($\Delta_{Xe}$) and in the framework of response-function formalism ($\Delta^{NR}_{Xe}$).

When the LDCP magnetic moment $\mu_{\chi}$ vanishes from Figs.~\ref{fig6.0} and \ref{fig7.0} it follows that
$\Delta_{Xe}(0)\simeq 0.17-0.18$ and $\Delta_{Ar}(0)\simeq 0.07-0.08$ for $m_{\chi}=200-1000\,\mbox{GeV}$.
As one can expect these quantities grow with increasing of $\mu_{\chi}$. Taking into account the
experimental limits on $\mu_{\chi}$ for different masses of the LDCP, from Fig.~\ref{fig6.0}a it is easy to see
that in general for $f\lesssim 10\,\mbox{TeV}$ and $m_{\chi}=200-500\,\mbox{GeV}$ the growth of $\Delta_{Xe}(\mu_{\chi})$
is rather weak. Such growth becomes more noticeable for $f\simeq 10\,\mbox{TeV}$ and $m_{\chi}=1\,\mbox{TeV}$
as well as for $f\simeq 15\,\mbox{TeV}$ especially in the part of the parameter space in which the partial cancellation of
the SM--like Higgs and $Z$ boson contributions to the LDCP--Xe scattering amplitude occurs, i.e. near $c_V^{\chi}\simeq 0.5$ and
$\varepsilon_H=-0.1$. For instance, for $f\simeq 10\,\mbox{TeV}$, $m_{\chi}=1\,\mbox{TeV}$, $c_V^{\chi}=0.5$ and
$\varepsilon_{H}=0$ the value of $\Delta_{Xe}\simeq 0.21$. For $f\simeq 15\,\mbox{TeV}$ and the same values of $c_V^{\chi}$ and
$\varepsilon_{H}$ the parameter $\Delta_{Xe}$ varies from $0.23$ to $0.3$ when $m_{\chi}$ changes from $200\,\mbox{GeV}$
to $1\,\mbox{TeV}$. At the same time if $c_V^{\chi}\simeq 0.5$ and $\varepsilon_H=-0.1$ the parameter $\Delta_{Xe}$
attains $0.23-0.29$ for $f\simeq 10\,\mbox{TeV}$ and $0.42-0.46$ for $f\simeq 15\,\mbox{TeV}$.
In the case of the LDCP--Ar scattering the relative growth of $\Delta_{Ar}(\mu_{\chi})$ is always
considerably larger as compared with $\Delta_{Xe}(\mu_{\chi})$. As mentioned before this is because $A_{Ar}\ll A_{Xe}$.
The values of $\Delta_{Xe}$ and $\Delta_{Ar}$ grow with increasing of the LDCP mass since the upper experimental
bound on the magnetic moment of dark matter particles gets stronger if $m_{\chi}$ decreases.

The enhancement of the event rate at low recoil energies might be caused not only by the
the magnetic moment of the LDCP but also non--zero electric charge $Q_{\chi}$ of the dark matter states.
Although in the CHMs $Q_{\chi}=0$ there are some other models in which dark matter particles with spin--$1/2$
carry very small fraction of the electron charge (millicharged DM) \cite{Goldberg:1986nk}. In order to distinguish the LDCP
from the millicharged DM and dark matter particles with $Q_{\chi}=0$ and $\mu_{\chi}=0$ it can be useful
to consider the function
\begin{equation}
\mathcal{Q}^A_R(\varepsilon_{rec})= \varepsilon_{rec} \mathcal{G}_A(\varepsilon_{rec})\,.
\label{40}
\end{equation}
If dark matter states have rather small (or zero) electric charge and/or very small (or zero) magnetic moment
the function $\mathcal{Q}^A_R(\varepsilon_{rec})$ diminishes considerably with decreasing $\varepsilon_{rec}$ at
low recoil energies. When spin--$1/2$ DM particles possess a sizeable magnetic moment while their electric charge
$Q_{\chi}$ is either too small or zero, $\mathcal{Q}^A_R(\varepsilon_{rec})$ goes to some finite value at
very low $\varepsilon_{rec}$. At the same time the presence of the dark matter states with appreciable electric
charge should lead to the growth of $\mathcal{Q}^A_R(\varepsilon_{rec})$ at low recoil energies because
the corresponding differential cross section contains term which is inversely proportional to $\varepsilon_{rec}^2$.

\begin{figure}[htbp]
\centering
\begin{subfigure}[b]{0.47\textwidth}
\centering
\includegraphics[width=\textwidth]{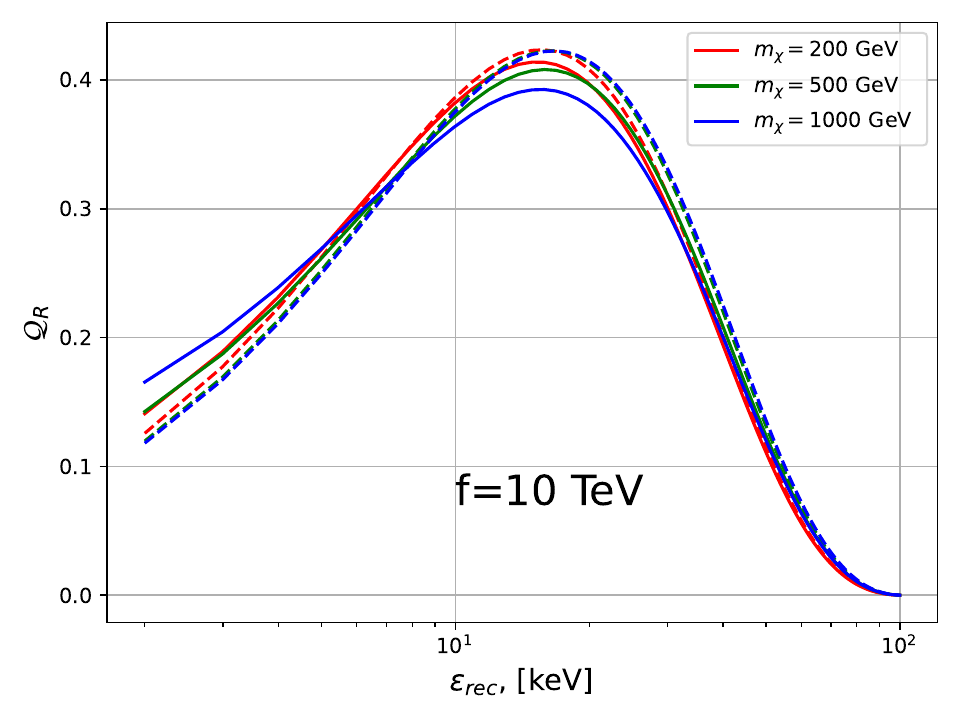}
\caption{}		
\end{subfigure}
\hfill  
\begin{subfigure}[b]{0.47\textwidth}
\centering
\includegraphics[width=\textwidth]{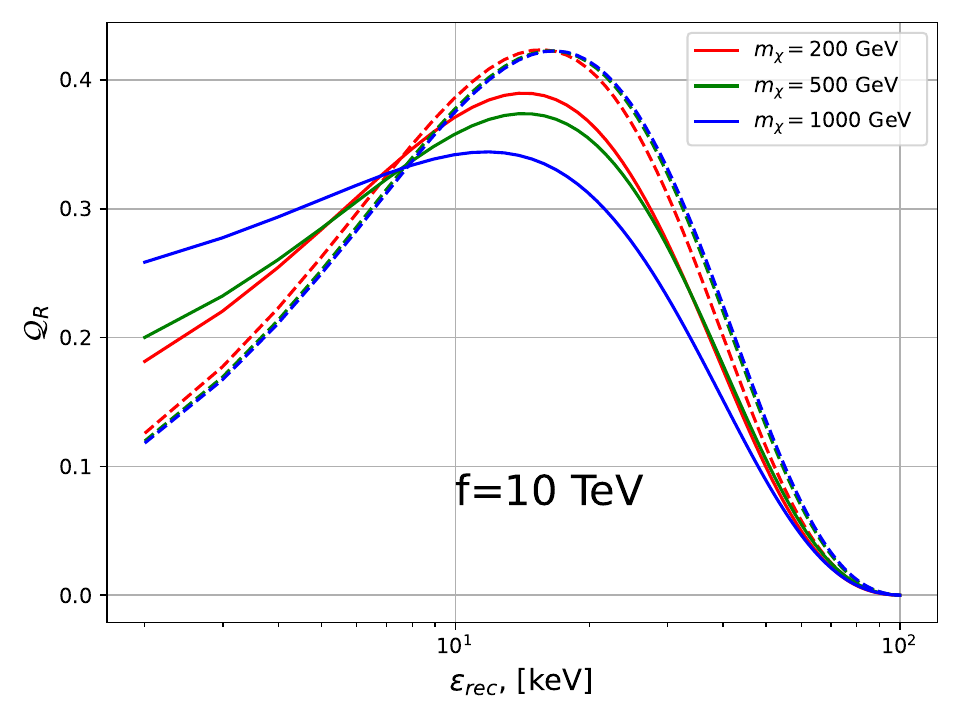}
\caption{}		
\end{subfigure}
\vspace{0.2cm}
\begin{subfigure}[b]{0.47\textwidth}
\centering
\includegraphics[width=\textwidth]{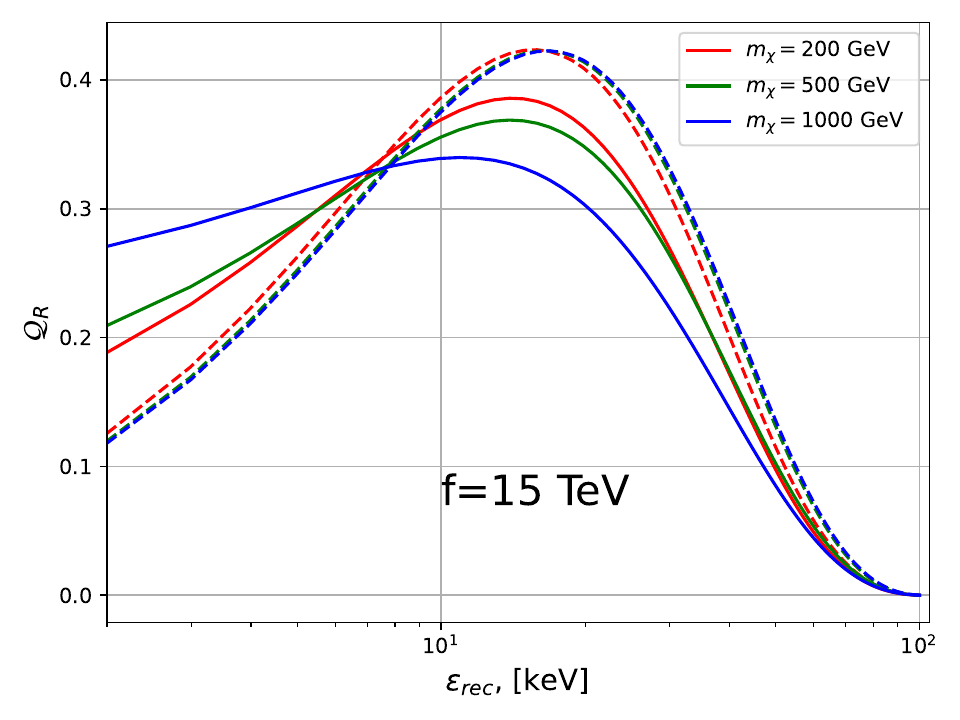}
\caption{}
\end{subfigure}
\hfill
\begin{subfigure}[b]{0.47\textwidth}
\centering
\includegraphics[width=\textwidth]{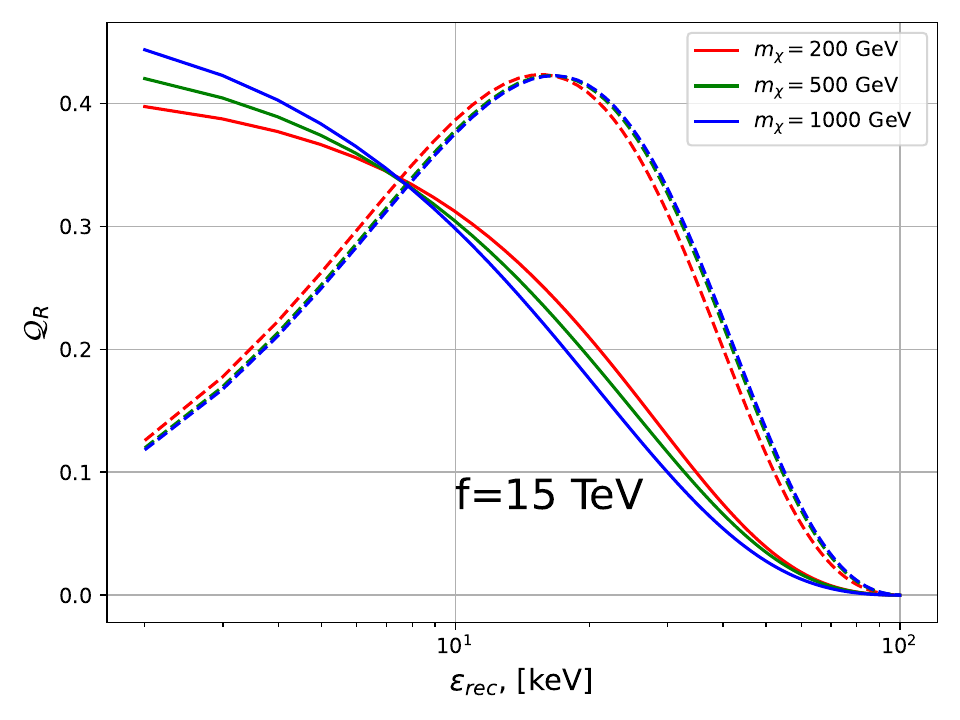}
\caption{}		
\end{subfigure}
\caption{Dependence of $\mathcal{Q}^{Xe}_R(\varepsilon_{rec})=\dfrac{\varepsilon_{rec}}{R^{Xe}_{tot}}\dfrac{dR_{Xe}}{d\varepsilon_{rec}}$
on recoil energy of Xe nuclei for
$f= 10\,\text{TeV}$, $c_V^{\chi}=0.5$ and $\varepsilon_{H}=0$ {\it (a)},
$f= 10\,\text{TeV}$, $c_V^{\chi}=0.5$ and $\varepsilon_{H}=-0.1$ {\it (b)},
$f= 15\,\text{TeV}$, $c_V^{\chi}=0.5$ and $\varepsilon_{H}=0$ {\it (c)} as well as
$f= 15\,\text{TeV}$, $c_V^{\chi}=0.5$ and $\varepsilon_{H}=-0.1$ {\it (d)}.
Red, green and blue lines are associated with $m_{\chi}=200\,\mbox{GeV}$,
$m_{\chi}=500\,\mbox{GeV}$ and $m_{\chi}=1000\,\mbox{GeV}$, respectively.
Solid and dashed lines correspond to $\mu_{\chi}=3\cdot \mu_{\chi}^{\text{exp}}(m_{\chi})$ and
$\mu_{\chi}=0$.}
\label{fig8.0}
\end{figure}

\begin{figure}[htbp]
\centering
\begin{subfigure}[b]{0.47\textwidth}
\centering
\includegraphics[width=\textwidth]{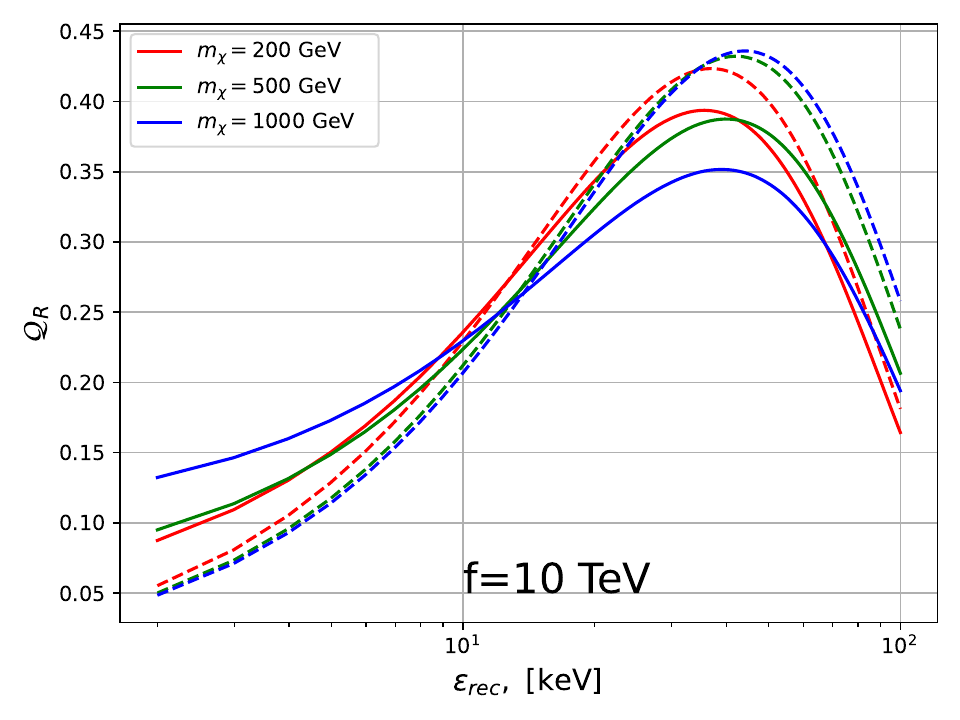}
\caption{}			
\end{subfigure}
\hfill  
\begin{subfigure}[b]{0.47\textwidth}
\centering
\includegraphics[width=\textwidth]{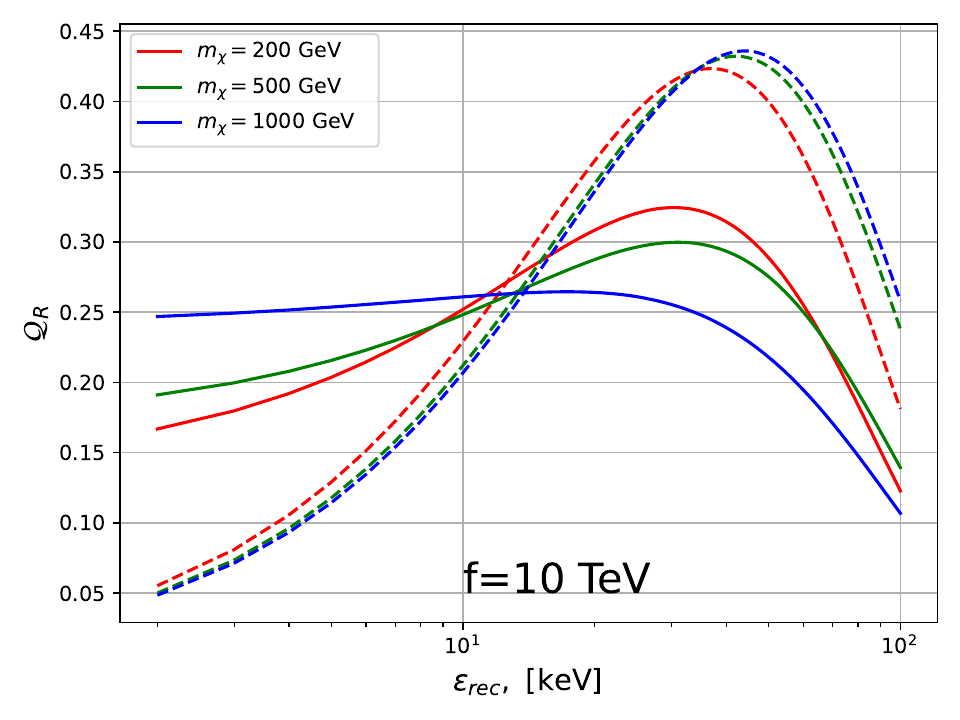}
\caption{}			
\end{subfigure}
\vspace{0.2cm}
\begin{subfigure}[b]{0.47\textwidth}
\centering
\includegraphics[width=\textwidth]{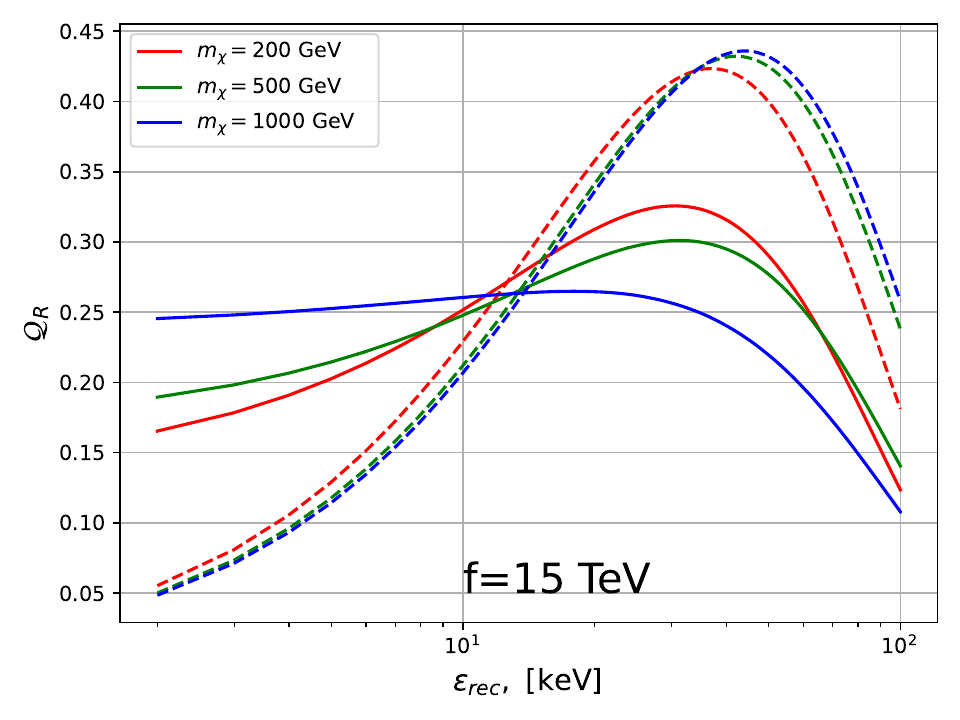}
\caption{}			
\end{subfigure}
\hfill
\begin{subfigure}[b]{0.47\textwidth}
\centering
\includegraphics[width=\textwidth]{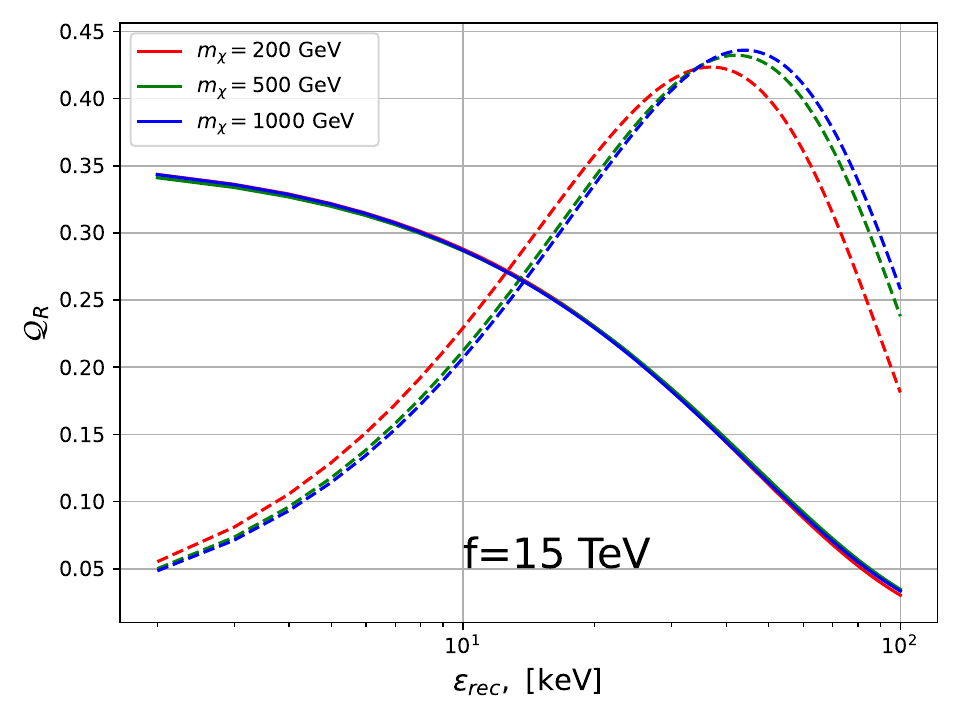}
\caption{}			
\end{subfigure}
\caption{Dependence of $\mathcal{Q}^{Ar}_R(\varepsilon_{rec})=\dfrac{\varepsilon_{rec}}{R^{Ar}_{tot}}\dfrac{dR_{Ar}}{d\varepsilon_{rec}}$
on recoil energy of Ar nuclei for
$f= 10\,\text{TeV}$, $c_V^{\chi}=0.5$ and $\varepsilon_{H}=0$ {\it (a)},
$f= 10\,\text{TeV}$, $c_V^{\chi}=0.5$ and $\varepsilon_{H}=-0.1$ {\it (b)},
$f= 15\,\text{TeV}$, $c_V^{\chi}=0.5$ and $\varepsilon_{H}=0$ {\it (c)} as well as
$f= 15\,\text{TeV}$, $c_V^{\chi}=0.5$ and $\varepsilon_{H}=-0.1$ {\it (d)}.
Red, green and blue lines are associated with $m_{\chi}=200\,\mbox{GeV}$,
$m_{\chi}=500\,\mbox{GeV}$ and $m_{\chi}=1000\,\mbox{GeV}$, respectively.
Solid and dashed lines correspond to $\mu_{\chi}=3\cdot \mu_{\chi}^{\text{exp}}(m_{\chi})$ and
$\mu_{\chi}=0$.}
\label{fig9.0}
\end{figure}

To illustrate this we plot $\mathcal{Q}^{Xe}_R(\varepsilon_{rec})$ and $\mathcal{Q}^{Ar}_R(\varepsilon_{rec})$
as functions of $\varepsilon_{rec}$ for $\mu_{\chi}=0$ and $\mu_{\chi}\simeq 3\cdot \mu^{exp}_{\chi}(m_{\chi})$
as well as for the LDCP masses $m_{\chi}=200\,\mbox{GeV}$, $500\,\mbox{GeV}$ and $1\,\mbox{TeV}$
in Figs.~\ref{fig8.0} and \ref{fig9.0}. To simplify our analysis we still keep $Q_{\chi}=0$.
From Fig.~\ref{fig8.0}a one can see that in general for $f\simeq 10\,\mbox{TeV}$ the functions $\mathcal{Q}^{Xe}_R(\varepsilon_{rec})$
computed for $\mu_{\chi}=0$ and $\mu_{\chi}\simeq 3\cdot \mu^{exp}_{\chi}(m_{\chi})$ remain close to
each other since the contribution to the LDCP--Xe scattering amplitude caused by the electromagnetic interaction is
relatively small. In the case of the LDCP--Ar scattering the differences between functions $\mathcal{Q}^{Ar}_R(\varepsilon_{rec})$
calculated for the values of the LDCP magnetic moment mentioned above become more noticeable (see Fig.~\ref{fig9.0}a).
Still it seems to be quite problematic to distinguish the scenarios with zero and non--zero LDCP magnetic moment
for $f\simeq 10\,\mbox{TeV}$ in the most part of the parameter space. However near $c_V^{\chi}\simeq 0.5$ and $\varepsilon_H=-0.1$, where
there is a partial cancellation of the SM--like Higgs and $Z$ boson contributions to the LDCP--Xe and LDCP--Ar
scattering amplitudes, the dependence of the functions $\mathcal{Q}^{Xe}_R(\varepsilon_{rec})$ and $\mathcal{Q}^{Ar}_R(\varepsilon_{rec})$
on $\varepsilon_{rec}$ for $\mu_{\chi}=0$ and $\mu_{\chi}\simeq 3\cdot \mu^{exp}_{\chi}(m_{\chi})$ tends to be rather different
especially for $m_{\chi}\simeq 1\,\mbox{TeV}$ (see Figs.~\ref{fig8.0}b and \ref{fig9.0}b).
The results presented in Figs.~\ref{fig8.0}c, \ref{fig8.0}d, \ref{fig9.0}c and \ref{fig9.0}d exhibit similar pattern
to the ones showed Figs.~\ref{fig8.0}b and \ref{fig9.0}b, i.e for $\mu_{\chi}\simeq 3\cdot \mu^{exp}_{\chi}(m_{\chi})$
the functions $\mathcal{Q}^{Xe}_R(\varepsilon_{rec})$ and $\mathcal{Q}^{Ar}_R(\varepsilon_{rec})$ do not decrease dramatically,
when $\varepsilon_{rec}\to 2\,\mbox{KeV}$, but go to some finite values in contrast to the scenarios with $\mu_{\chi}=0$.
Thus the investigation of the dependence of the function $\mathcal{Q}^A_R(\varepsilon_{rec})$ on the recoil energy may
provide an important complementary information regarding the electromagnetic properties of the dark matter states.

\section{Conclusion}

A few years ago PICO experiment placed stringent limits on the dark matter (DM) magnetic moment. More recently xenon based experiment LZ
significantly strengthened bounds on the spin--independent dark matter--nucleon cross section. Since within last few years
LZ experiment improved these upper bounds by an order of magnitude one may hope that the DM signal can be detected in the near future.
Motivated by the substantial progress in the sensitivity of the direct detection experiments in this article we examined the
interactions of the Dirac DM fermions with nucleons as well as with Xe and Ar nuclei within the composite Higgs models (CHMs).

In the CHMs the strongly coupled sector may result in two SM singlet Weyl fermions which form the neutral Dirac state $\chi$.
The couplings of such lightest Dirac composite particle (LDCP) to all SM fields can be quite suppressed. Moreover such LDCP may be
stable composing the DM in our Universe. This scenario is realised in the $E_6$ inspired composite Higgs model (E$_6$CHM) in which
the strongly interacting sector possesses the approximate $SU(6)$ symmetry. The breakdown of $SU(6)$ near the compositeness scale
$f\gtrsim 5\,\mbox{TeV}$ down to $SU(5)$, that involves the SM gauge group, gives rise to a set of the pNGB states.
These states form the SM Higgs doublet $H$, the scalar colour triplet $T$ and the SM singlet pseudoscalar $\phi_0$ which tend
to be considerably lighter than $f$. If all composite fermions in the E$_6$CHM gain masses of the order of $f$ the baryon number
conservation ensures that the colour triplet $T$ is stable. Such scenarios have been already excluded.
Nevertheless the approximate $U(1)_E$ symmetry can lead to the suppression of the mass of the LDCP that carries baryon numbers $1/3$
in the E$_6$CHM. When the LDCP is lighter than the scalar colour triplet, $T$ decays into $\bar{\chi}$ and $b$--quark while
$\chi$ remains stable. The presence of the scalar colored triplet with mass $\lesssim 2\,\mbox{TeV}$ in the particle spectrum
is a distinctive feature of the $E_6$CHM. At the LHC these states can be pair--produced leading to some enhancement of the cross section
$pp\to b\overline{b} + E^{\rm miss}_{T} + X$, where $E^{\rm miss}_{T}$ is the energy and momentum carried away by the LDCP.

It is expected that in the CHMs the LDCP has a magnetic dipole moment $\mu_{\chi}\sim e/f$ while
the DM electric dipole moment vanishes if CP in the strongly interacting sector is preserved. Then the current experimental constraints
on the DM magnetic moment imply that the compositeness scale $f$ should be larger than $10^{4}\,\mbox{TeV}$. The CHMs with so high
scale $f$ are strongly disfavoured because of the enormous degree of tuning which is required to get the $125\,\mbox{GeV}$ Higgs state.
Nonetheless the DM magnetic dipole moment can be substantially smaller than $e/f$ if all constituents, that compose the LDCP, have no
electric charges. In the E$_6$CHM $\mu_{\chi}$ can be further suppressed by the approximate $U(1)_E$ symmetry so that the
experimental constraints on $\mu_{\chi}$ may be fulfilled even for $f\sim 10\,\mbox{TeV}$. Therefore in this article we restricted
our consideration to the CHMs with approximate $U(1)$ symmetry. This symmetry permits to suppress the LDCP mass $m_{\chi}$,
its magnetic moment $\mu_{\chi}$ and the dimensionless coupling of the LDCP to the Higgs doublet $\varepsilon_{H}$.

To avoid too large degree of fine--tuning in our analysis we varied the compositeness scale $f$ from $5\,\mbox{TeV}$ to $15\,\mbox{TeV}$.
We also assumed that $m_{\chi}$ and the dimensionless coupling $\varepsilon_{H}$, which violate global $U(1)$ symmetry, are
quite small, i.e. $m_{\chi}\le 1\,\mbox{TeV}$ and $|\varepsilon_{H}|\le 0.1$. The LDCP mass is taken to be larger than $200\,\mbox{GeV}$
to allow for efficient annihilation $\chi\bar{\chi}\to t\bar{t}$. In the E$_6$CHM such annihilation process can be efficient if $m_{\chi}$ is
rather close to half the mass of the SM singlet pseudoscalar $\phi_0$.

Within the approximation mentioned above the interaction of the LDCP with nucleons is dominated by the $t$--channel exchange of the $Z$ boson.
The vector and axial--vector couplings of the $Z$ boson to the LDCP are inversely proportional to $f^2$ and defined by the dimensionless
couplings $c^{\chi}_V$ and $c^{\chi}_{PV}$ respectively. Both couplings $c^{\chi}_V$ and $c^{\chi}_{PV}$ are expected to be of order of
unity. The spin--independent (SI) DM--nucleon scattering cross section is mostly determined by $f$ and $c^{\chi}_V$.
When $f\simeq 5\,\mbox{TeV}$ and $c^{\chi}_V\sim 1$ this cross section is typically considerably larger than
the corresponding experimental bounds. Hence in such scenarios the LDCP can form only a small fraction $\xi$
of the total DM relic abundance. In our analysis $\xi$ and $c^{\chi}_V$ are required to be larger than $0.1$ and $0.5$ respectively\,.
If $f\gtrsim 2\,\mbox{TeV}$ the spin-dependent (SD) LDCP--nucleon scattering cross sections are substantially smaller than the experimental bounds.
For $\xi\ge 0.1$ we identified the parts of the parameter space, which are consistent with the SI direct detection limits, and specified
a set of benchmark scenarios.

Using the benchmark scenarios we explored the LDCP--Xe and LDCP--Ar elastic scattering. Our analysis indicated that for
non--zero LDCP magnetic dipole moment the differential cross section of the DM--nucleus scattering could grow much stronger with decreasing of
nuclear recoil energy $\varepsilon_{rec}$ as compared with the scenarios with $\mu_{\chi}=0$. In the direct detection experiments
this may allow to distinguish the DM fermions with $\mu_{\chi}\ne 0$ from other types of dark matter states which don't have
magnetic moment. In this context we examined the dependence of the differential cross section and the normalized differential event rate
$\mathcal{G}_A(\varepsilon_{rec})$ of the LDCP--Xe and LDCP--Ar scattering on $\varepsilon_{rec}$. For $f \lesssim 10\,\mbox{TeV}$
the contributions of the electromagnetic interaction to the LDCP--Xe and LDCP--Ar scattering amplitudes in the CHMs are quite small because of the
stringent experimental limits on $\mu_{\chi}$. Therefore it is going to be rather problematic to probe the electromagnetic properties of
the DM fermions if $f \lesssim 10\,\mbox{TeV}$. The $Z$ boson and Higgs contributions to the amplitude of the LDCP–-nucleus scattering
diminish with increasing $f$. As a consequence for $f \simeq 15\,\text{TeV}$ the electromagnetic contribution can induce a sizable enhancement
of the normalized differential event rate at recoil energies $\varepsilon_{\text{rec}} \lesssim 10\,\text{keV}$. This enhancement, however,
occurs only if $c_V^{\chi}$ is considerably smaller than unity. Furthermore the growth of $\mathcal{G}_A(\varepsilon_{rec})$ at low recoil
energies is getting more substantial in the part of the CHM parameter space near $c_V^{\chi}\simeq 0.5$ and $\varepsilon_H=-0.1$,
where the SM--like Higgs and $Z$ boson contributions to the LDCP--nucleus scattering amplitude partially cancel each other.
From Tables~\ref{tab:BM} and \ref{tab:fractions} it follows that the electromagnetic interaction may result in the noticeable enhancement of
$\mathcal{G}_A(\varepsilon_{rec})$ at $\varepsilon_{rec}\lesssim 5\,\mbox{KeV}$ only when $\sigma^N_{SI}$ is of order of a few $\mbox{yb}$
or even smaller. Such enhancement is bigger in the case of argon nuclei since $A_{Ar}\ll A_{Xe}$.

To quantify the low energy enhancement of $\mathcal{G}_A(\varepsilon_{\text{rec}})$ numerically
we estimated the relative fraction of the events with $\varepsilon_{rec}$ in the range $2-5\,\mbox{KeV}$, i.e. $\Delta_{A}(\mu_{\chi})$.
When the LDCP magnetic moment $\mu_{\chi}$ vanishes, $\Delta_{A}(\mu_{\chi})$ attain their minimal values:
$\Delta_{\mathrm{Xe}}(0)\simeq 0.17$--$0.18$ and $\Delta_{\mathrm{Ar}}(0)\simeq 0.07$--$0.08$ for $m_{\chi}=200$--$1000\,\mathrm{GeV}$.
As expected $\Delta_{A}(\mu_{\chi})$ grow with increasing $\mu_{\chi}$. However they remain bounded by $\Delta_{\mathrm{Xe}}(\mu_{\chi})\simeq 0.47$
and $\Delta_{\mathrm{Ar}}(\mu_{\chi})\simeq 0.38$, which correspond to the purely magnetic dark matter limit.
Although $\Delta_{A}(\mu_{\chi})$ increases with $\mu_{\chi}$ such growth remains rather constrained for $f\lesssim 10\,\text{TeV}$
throughout the major part of the phenomenologically viable parameter space. The low energy enhancement becomes more noticeable
for $f\simeq 15\,\mbox{TeV}$ when $c_V^{\chi}$ is substantially smaller than unity.
For $f\gtrsim 10\,\mbox{TeV}$ the quantities $\Delta_{A}(\mu_{\chi})$ approach their maximal values near $c_V^{\chi}=0.5$ and
$\varepsilon_{H}=-0.1$\,. In particular, for $f\simeq 15\,\mbox{TeV}$ and $m_{\chi}=200-1000\,\mbox{GeV}$ the quantities
$\Delta_{Xe}(\mu_{\chi})$ and $\Delta_{Ar}(\mu_{\chi})$ can reach $0.42-0.46$ and $0.36-0.37$ respectively.
A sufficiently big difference between $\Delta_{A}(0)$ and $\Delta_{A}(\mu_{\chi})$, which is attained for $\sigma^N_{SI}$ of less than
a few $\mbox{yb}$, should permit to detect the enhancement of the event rate at low $\varepsilon_{rec}$ caused by the electromagnetic interaction.
The corresponding intervals of variations enlarge with increasing $m_{\chi}$ because the upper experimental limit
on the magnetic moment of the DM fermions is weaker for larger $m_{\chi}$.

The growth of the event rate at low recoil energies may not be caused by only non--zero magnetic moment
of the DM states. In this connection it is useful to study the dependence of function $\mathcal{Q}^A_R(\varepsilon_{rec})=\varepsilon_{rec}\mathcal{G}_A(\varepsilon_{rec})$ on $\varepsilon_{rec}$.
If DM is predominantly composed of the states which carry small but still appreciable fraction of the electron charge $Q_{\chi}$ then
$\mathcal{Q}^A_R(\varepsilon_{rec})$ has to grow with decreasing $\varepsilon_{rec}$ at low recoil energies.
When $Q_{\chi}$ and $\mu_{\chi}$ are negligibly small $\mathcal{Q}^A_R(\varepsilon_{rec})$ should
diminish rapidly with decreasing $\varepsilon_{rec}$ at low recoil energies. Here we argued that
in the scenarios with $\sigma^N_{SI}$ of less than a few $\mbox{yb}$ the functions $\mathcal{Q}^{Xe}_R(\varepsilon_{rec})$ and $\mathcal{Q}^{Ar}_R(\varepsilon_{rec})$ do not decrease dramatically in the limit $\varepsilon_{rec}\to 2\,\mbox{KeV}$,
but approach some finite values. Therefore the analysis of the function $\mathcal{Q}^A_R(\varepsilon_{rec})$
may shed light on the electromagnetic properties of the DM states. The results of our investigations also
indicate that the argon based experiments, such as DarkSide project, are better suited for exploration of
the electromagnetic properties of the DM particles.

\section*{Acknowledgements}

\vspace{-2mm}
R. N. acknowledges fruitful discussions with A.~O.~Barvinsky, S.~V.~Demidov, D.~S.~Gorbunov, D.~G.~Levkov,
K.~V.~Stepanyantz and M.~I.~Vysotsky.

\end{document}